\documentclass[aps,prd,nofootinbib,superscriptaddress,prd,aps,floatfix]{revtex4}
\usepackage[latin9]{inputenc}
\usepackage{diagbox} 
\usepackage{bm}
\usepackage{color,textcomp,amsmath,amssymb,graphicx,amsfonts,graphics,epstopdf,slashed,multirow,adjustbox,lipsum,rotating,xcolor,wrapfig,epsfig}
\usepackage[unicode=true, bookmarks=false, breaklinks=false,pdfborder={0 0 1},backref=section,colorlinks=false]{hyperref}
\hypersetup{colorlinks,bookmarksopen,bookmarksnumbered,linkcolor=blu,pdfstartview=FitH,urlcolor=rossos,citecolor=rossos}
\usepackage[justification=raggedright, margin=5mm,font={sf,small},labelfont=bf, format=plain]{caption}

\definecolor{rosso}{cmyk}{0,1,1,0.4}
\definecolor{rossos}{cmyk}{0,1,1,0.55}
\definecolor{rossoc}{cmyk}{0,1,1,0.2}
\definecolor{blu}{cmyk}{1,1,0,0.3}
\definecolor{blus}{cmyk}{1,1,0,0.6}
\definecolor{bluc}{cmyk}{1,1,0,0.1}
\definecolor{verde}{cmyk}{0.92,0,0.59,0.25}
\definecolor{verdec}{cmyk}{0.92,0,0.59,0.15}
\definecolor{verdes}{cmyk}{0.92,0,0.59,0.4}

\begin{document}

\title{ Probing the Dark Matter of Three-loop Radiative Neutrino Mass Generation Model with the Cherenkov Telescope Array}

\author{Talal Ahmed Chowdhury}
\email{talal@du.ac.bd}
\affiliation{Department of Physics, University of Dhaka, P.O. Box 1000,
Dhaka, Bangladesh.}
\affiliation{The Abdus Salam International Centre for Theoretical Physics, Strada
Costiera 11, I-34014, Trieste, Italy.}

\author{Saquib Hassan}
\email{saquib.hassan@physics.ox.ac.uk}
\affiliation{Rudolf Peierls Centre for Theoretical Physics, University of Oxford, Oxford, OX1 3PU, UK}

\author{Jahid Hossain}
\email{jhossain2@huskers.unl.edu}
\affiliation{Department of Physics and Astronomy, University of Nebraska-Lincoln, Lincoln, NE 68588-0299, USA}

\author{Salah Nasri}
\email{snasri@uaeu.ac.ae}
\affiliation{Department of Physics, UAE University, P.O. Box
17551, Al-Ain, United Arab Emirates}
\affiliation{The Abdus Salam International Centre for Theoretical Physics, Strada
Costiera 11, I-34014, Trieste, Italy.}

\author{Mahmud Ashraf Shamim}
\email{mahmud.ashamim84@gmail.com}
\affiliation{Department of Physics, DPS STS School Dhaka, P. O. Box 1230, Dhaka, Bangladesh}

\begin{abstract}
We investigate the prospect of detecting the Dark Matter (DM) candidate in the three-loop radiative neutrino mass generation model extended with large electroweak multiplets of the Standard Model (SM) gauge group, at the future imaging atmospheric Cherenkov telescope known as the Cherenkov Telescope Array (CTA). We find that the addition of such large electroweak multiplets leads to a sizable Sommerfeld enhanced annihilation of the DM with $O(\text{TeV})$ mass, into the SM gauge bosons which results in continuum and line-like spectra of very high energy (VHE) gamma-rays, and therefore becomes observable for the CTA. We determine the viable models by setting the upper limit on the $SU(2)_{L}$ isospin of the multiplets from the partial-wave unitarity constraints and the appearance of low-scale Landau pole in the gauge coupling. Afterwards, by considering the continuum VHE gamma-rays produced from the DM annihilation at the Galactic center, we probe the parameter space of the model using the sensitivity reach of the CTA.
\end{abstract}

\maketitle
\section{Introduction}\label{intro}

In recent years, the Imaging Atmospheric Cherenkov Telescopes (IACT)
have not only opened new avenues for the ground-based very high energy gamma-ray astronomy \cite{Rieger:2013rwa, DeAngelis:2018lra, DiSciascio:2019lse} but also offered a testing ground for the Dark Matter (DM) of the universe \cite{Bergstrom:1997fj, Bringmann:2012ez}. The parameter space of the DM which has the electroweak quantum numbers and mass in the GeV-TeV range has been probed by currently operating major IACTs: High Energy Stereoscopic System (H.E.S.S.) \cite{Rinchiuso:2019rrh, Abdallah:2016ygi, Abdallah:2018qtu, Abramowski:2013ax, Abdalla:2016olq}, Major Atmospheric Gamma Imaging Cherenkov Telescope (MAGIC) \cite{Paredes:2020uvz, Doro:2017dqn} and Very Energetic Radiation Imaging Telescope Array (VERITAS) \cite{Zitzer:2017xlo}. The Cherenkov Telescope Array (CTA) which is an ongoing international development project for the next-generation IACT, will have the capability to observe gamma-ray of energy from 20 GeV to at least 300 TeV over a large area and wide range of view (up to $10^{\circ}$) with more than 100 telescopes located in northern and southern hemispheres. It will allow CTA to achieve a sensitivity about a factor 10 better than current instruments such as H.E.S.S., MAGIC or VERITAS \cite{Actis_2011, Doro:2012xx, bsacharya, Pierre:2014tra, Silverwood:2014yza, Roszkowski:2014iqa, Lefranc:2015pza, Ibarra:2015tya, Lefranc:2016dgx, Lefranc:2016fgn, Bigongiari:2016amk, Conrad:2016jww, Balazs:2017hxh, CTAConsortium:2018tzg, Hryczuk:2019nql, Acharyya:2020sbj, Rinchiuso:2020skh}. The DM charged under the Standard Model (SM) gauge group and with the mass in the multi-TeV range can produce highly energetic diffuse or monochromatic gamma-rays (based on the annihilation channels) which will be within the observational reach of the CTA. Therefore in this paper, we have explored the detection possibility of the TeV DM candidate of the Three-loop Radiative Neutrino Mass Model, known as the Krauss-Nasri-Trodden model, at the CTA.

The Krauss-Nasri-Trodden (KNT) model \cite{Krauss:2002px} is one of the early models of radiative neutrino mass generation (for a comprehensive review, please see \cite{Cai:2017jrq}) which ties the origin and smallness of the neutrino mass with the DM of the universe. The additional Beyond Standard Model (BSM) fields of the model are two single charged singlet scalars, $S^{+}_{1},\,S^{+}_{2}$ and three SM singlet RH neutrinos, $N_{R_{i}},\,i=1,2,3$ with masses lie in the GeV-TeV range. There is a $Z_{2}$ symmetry, $\{S^{+}_{2},N_{R_{i}}\}\rightarrow \{-S^{+}_{2},-N_{R_{i}}\}$ which not only omits the tree-level Dirac mass term, $\overline{L_{L}}N_{R_{i}}H$, where $L_{L}$ and $H$ are the SM left-handed lepton doublet and Higgs doublet respectively, but also stabilizes the lightest singlet RH neutrino $N_{R_{1}}$ to play the role of DM. Besides, the KNT model can be generalized \cite{Chen:2014ska} by replacing $S^{+}_{2}$ with $\mathbf{\Phi}$ having integer isospin and hypercharge, $Y=1$ and $N_{R_{i}}$ with $\mathbf{F}_{i}$ that has integer isospin and $Y=0$ under SM gauge group which leaves the three-loop neutrino mass topology invariant. In the generalized KNT model, the lightest neutral fermion component, $F_{1}^{0}$ is the viable DM candidate. Such replacements in KNT model with large electroweak multiplets have been studied for triplet \cite{Ahriche:2014cda}, 5-plet \cite{Ahriche:2014oda} and 7-plet \cite{Ahriche:2015wha} cases. In \cite{Chowdhury:2018mrr}, it has been  shown that for most of the DM mass range from 1 to 50 TeV, the Sommerfeld enhanced annihilation cross-section to SM gauge bosons in 5-plet and 7-plet are already within reach of H.E.S.S. and the future CTA because more charged component fields of the fermion multiplet contribute to the enhancement and thus lead to a larger annihilation cross-section. Therefore, to probe viable DM candidate in the class of generalized KNT model, we are left with only the minimal model with singlet fermion and next-to-minimal triplet model where DM is the neutral component of a fermion triplet.

The DM of the minimal KNT model, being the singlet under the SM gauge group, has different DM parameter space \cite{Ahriche:2013zwa} than the triplet or 5-plet and 7-plet models which have dominant gauge interactions controlling the viable DM parameter space. For this reason, in our CTA sensitivity study, we have addressed the triplet DM because it shares the same region of parameter space as the 5-plet and 7-plet, and left the singlet case for a separate analysis.

The article is organized as follows. In section \ref{model}, we present the model and calculate the possible upper bound on the $SU(2)_{L}$ isospin of the electroweak multiplets in the generalized KNT model. In section \ref{DMtriplet}, we describe the Sommerfeld enhanced DM annihilation cross-section, and the gamma-ray flux originated from the DM annihilation in the triplet KNT model. Section  \ref{ctasec} delineates the working principle of the CTA, its instrument response functions and possible sources of background, and lastly the observation region we consider in our study. In section    \ref{resultssec}, we outline our methods to determine the expected gamma-ray counts from the DM signal and backgrounds, and use the Likelihood analysis to set the upper limit on the DM annihilation cross-section in the KNT model. Finally, we conclude in section
 \ref{conclusionsec}. In appendix \ref{partialapp}, we present the detailed calculations relevant for our partial-wave unitarity constraints.

\section{The Model}\label{model}
The BSM fields in the generalized KNT model, charged under SM gauge group, $SU(3)_{c}\times SU(2)_{L}\times U(1)_{Y}$ are
\begin{equation}
\text{Complex scalars:}\,\,\,S^{+}_{1}\sim (0,0,1),\,\,\mathbf{\Phi}\sim (0,j_{\phi},1),\,\,\text{and}\,\,\text{Real fermions:}\,\,\mathbf{F}_{1,2,3}\sim (0,j_{F},0)\,,
\label{particle}
\end{equation}
where $j_{\phi}$ and $j_{F}$ are integer isospin of $SU(2)_{L}$. For example the triplet model would contain, $\mathbf{\Phi}\sim (0,1,1)$ and $\mathbf{F}_{1,2,3}\sim (0,1,0)$\footnote{We consider the electric charge to be $Q=T^{3}+Y$, where $T^{3}$ and $Y$ are the diagonal generators of $SU(2)_{L}$ and $U(1)_Y$, respectively.}.
 
The generalized KNT Lagrangian that has the additional $Z_{2}$ symmetry, is given by,
\begin{equation}
{\cal L}\supset {\cal L}_{SM}+\{f_{\alpha\beta} \overline{L^{c}_{\alpha}}.L_{\beta}S_{1}^{+}
+g_{i \alpha}\overline{\mathbf{F}_{i}}.\mathbf{\Phi}.e_{\alpha_{R}}+h.c\}-\frac{1}{2}\overline{\mathbf{F}^{c}_{i}}M_{F_{ij}}\mathbf{F}_{j}-V(H,\mathbf{\Phi},S_{1})+h.c\,,
\label{eq1}
\end{equation} 
where, c denotes the charge conjugation and dot sign, in shorthand, refers to appropriate $SU(2)$ contractions. Also $L_{\alpha}$ and $e_{R_{\alpha}}$ are the left-handed (LH) lepton doublet and right-handed (RH) charged leptons, respectively, and Greek alphabet $\alpha$ stands for the flavor index, i.e. $\alpha=e,\mu,\tau$. Moreover, $f_{\alpha\beta}$ and $g_{i\alpha}$ (where $i$ is the generation index of the fermionic multiplets, $i=1,2,3$) are the Yukawa couplings appearing in Eq.(\ref{eq1}) which can be considered as the components of a complex antisymmetric matrix, $F$ and a general complex matrix, $G$, respectively. $H$ is the SM Higgs doublet. Finally, $V(H,\mathbf{\Phi},S_{1})$ denotes the scalar potential. 

We have seen that the generalized KNT model contains $SU(2)_{L}$ fermionic and scalar multiplets of integer isospins. In the case of the minimal KNT model \cite{Krauss:2002px}, the yukawa term of Eq. (\ref{eq1}), $g_{i \alpha}\overline{\mathbf{F}_{i}}.\mathbf{\Phi}.e_{\alpha_{R}}$ boils down to the term containing three real Majorana fermions, $\mathbf{F}_{i}\equiv N_{i}$ which are SM singlet, and single-charged $SU(2)_{L}$ singlet scalar, $\mathbf{\Phi}\equiv S_{2}^{+}$. Therefore, in generalized KNT model, one would require three real $SU(2)_{L}$ fermionic multiplets, $\mathbf{F}_{i}$ to embed the minimal KNT model's fermions as one of the component fields in those multiplets. This requires the $\mathbf{F}_{i}$ to have integer isospin and zero hypercharge. Now the SM invariance of the yukawa term forces the $SU(2)_{L}$ scalar multiplet, $\mathbf{\Phi}$ to have equal integer isospin of the fermionic multiplets, i.e. $j_{\phi}=j_{F}$, and hypercharge, $Y=1$. As we can see, the single-charged scalar of the minimal KNT model is now a component field of the complex scalar multiplet.

The mass splittings in fermionic component fields are zero at tree-level and only receive $O(100\, \text{MeV})$ splittings due to the radiative correction after the electroweak symmetry breaking 
\cite{Cirelli:2005uq, Ibe:2012sx, McKay:2017xlc}. On the other hand, the mass splittings among component fields of the scalar multiplet $\mathbf{\Phi}$ is controlled by $\lambda_{H\phi 2}(\mathbf{\Phi}^{\dagger}.H).(H^\dagger.\mathbf{\Phi})$ term of the scalar potential after electroweak symmetry breaking and the splittings, $\Delta m$ allowed by the electroweak precision observables only lead to $\Delta m^{2}_{ij}/M_{0}^{2}\sim 10^{-3}$ if the invariant mass of the scalar multiplet is, $M_{0}=10$ TeV, and for $M_{0}>10$ TeV, the ratio becomes even smaller as shown in \cite{Chowdhury:2018nhd}. Therefore, the component fields of both scalar and fermion multiplets are almost degenerate at the TeV scale. Consequently, such near-degeneracy of fermion component fields does not diminish the Sommerfeld enhancement of DM annihilation.

As we have seen that the large isospin multiplets of the $SU(2)_{L}$ are allowed in the generalized KNT model, one could ask the upper bound on the electroweak quantum numbers for such large multiplets. In the following sections \ref{partialsec} and \ref{landaupolesec}, we have used the arguments from the Partial-wave unitarity and appearance of the low-scale Landau pole to set an upper limit on the isospin in the generalized KNT model.

\subsection{Partial-Wave unitarity constraints}\label{partialsec}
\begin{figure}[h!]
\vspace{-1.75cm}
\centerline{\includegraphics[width=15cm]{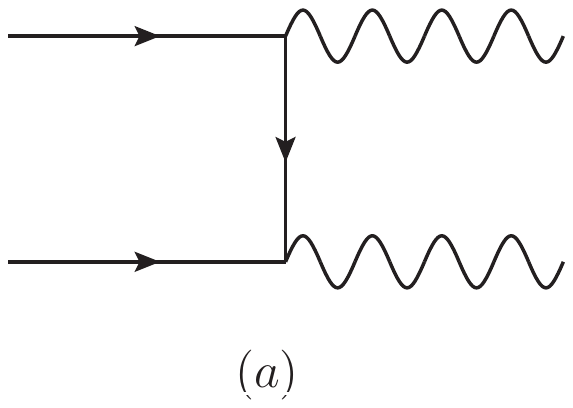}\hspace{-8.5cm}\includegraphics[width=15cm]{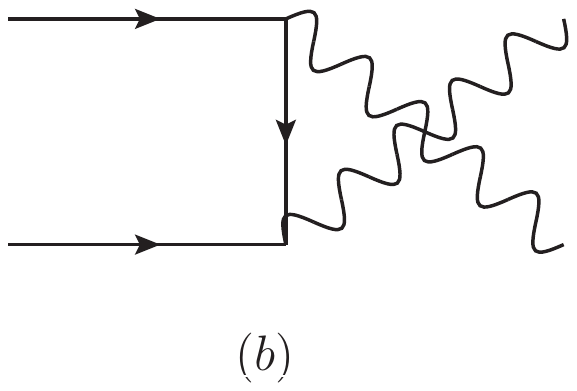}\hspace{-8.5cm}\includegraphics[width=15cm]{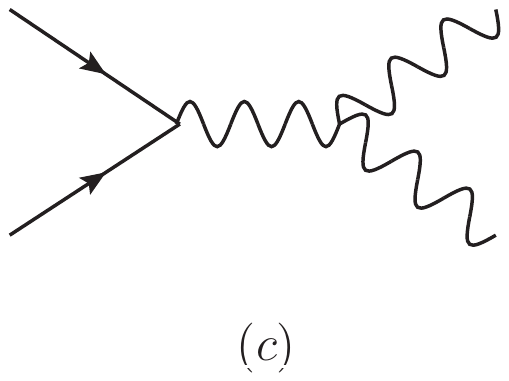}}
\vspace{-16.5cm}
\caption{Partial-wave unitarity constraints on scattering processes $FF\rightarrow VV$}
\label{partialfig}
\end{figure}

To set the partial-wave unitarity constraints on the electroweak quantum numbers of large multiplets in the KNT model, we have looked into the tree-level scattering of the component fields of the fermion multiplets into the SM gauge bosons, $FF\rightarrow VV$, i.e. into $W\,W,\,Z\,Z,\,\gamma\,\gamma,\,\gamma\,Z$, as shown in Fig. \ref{partialfig}. Similar analysis with SM gauge bosons as the final states was done in \cite{Babu:2011sd}. Moreover, the upper bound on the isospin for the scalar multiplet using same final states, was addressed by \cite{Hally:2012pu}.

Let us consider the $2\rightarrow 2 $ scattering amplitude in momentum space with initial and final 2-particle states $|i\rangle$ and $|f\rangle$, respectively, as
\begin{equation}
\langle f|T|i\rangle=(2\pi)^{4}\delta^{4}(P_{f}-P_{i}){\cal T}_{fi}(\sqrt{s},\cos\theta)\,,
\label{partial1}
\end{equation}
where, $s$ is the center-of-mass energy. Here $T$ captures the interaction part of the S-matrix, $S=1+i T$. From Eq. (\ref{partial1}) by using the Jacob-Wick expansion \cite{Jacob:1959at, Schuessler:2007av, DiLuzio:2016sur}, the corresponding partial-wave amplitude of total angular momentum $J$ is,
\begin{equation}
a^{J}_{f i}=\frac{\beta^{1/4}(s,m^{2}_{i_{1}},m^{2}_{i_{2}})\beta^{1/4}(s,m^{2}_{f_{1}},m^{2}_{f_{2}})}{32\pi s}\int^{1}_{-1}
d(\cos\theta)d^{J}_{\mu_{i}\mu_{f}}(\theta){\cal T}_{fi}(\sqrt{s},\cos\theta)\,,
\label{partial2}
\end{equation}
where, $d^{J}_{\mu_{i}\mu_{f}}$ is the $J$-th Wigner $d$-function with $\mu_{i}=\lambda_{i_{1}}-\lambda_{i_{2}}$ and $\mu_{f}=\lambda_{f_{1}}-\lambda_{f_{2}}$ defined in terms of the helicities of the initial $(\lambda_{i_{1}},\lambda_{i_{2}})$ and final $(\lambda_{f_{1}},\lambda_{f_{2}})$ states. In addition, $\beta(x,y,z)=x^{2}+y^{2}+z^{2}-2 x y- 2 y z -2 z x$ is the Kallen function. Moreover, a factor of $1/\sqrt{2}$ should be multiplied at the right hand side of Eq. (\ref{partial2}) for any identical pairs of particles either in the initial or final states.

The unitarity condition on the S-matrix, $S^{\dagger}S=1$, implies that
\begin{equation}
\frac{1}{2 i}(a^{J}_{fi}-a^{J*}_{if})\geq\sum_{n}a^{J*}_{fn}a^{J}_{ni}\,,\label{partial3}
\end{equation}
where the inequality arises because the sum over $n$ is restricted to 2-particle states only. If we consider elastic scattering for which, $f=i$, we have from Eq.(\ref{partial3}),
\begin{equation}
\text{Im}a^{J}_{ii}\geq |a^{J}_{ii}|^{2}+\sum_{k\neq i}|a^{J\,\text{inel}}_{ki}|^{2}\,,\label{partial4}
\end{equation}
where the sum over $k\neq i$ are taken over all of the possible 2-particle inelastic channels.
By writing, $|a^{J}|^{2}=(\text{Re}a^{J})^2+(\text{Im}a^{J})^2$, we have,
\begin{equation}
\text{Im}a^{J}_{ii}(1-\text{Im}a^{J}_{ii})\geq (\text{Re}a^{J}_{ii})^{2}+\sum_{k\neq i}|a^{J\,\text{inel}}_{ki}|^{2}\,.\label{partial5}
\end{equation}
Now that $a^{J}_{ii}$ lies on the unitary circle, the left-hand side of Eq.(\ref{partial5}) is bounded by 1/4, we can have in absence of any inelastic channel,
\begin{equation}
\text{Re}\,a^{J}_{ii}\leq \frac{1}{2}\,.
\label{partial6}
\end{equation}
Therefore, the possible maximal bound one can set on the real part of an inelastic channel $k$ is,
\begin{equation}
\text{Re}\,a^{J\,\text{inel}}_{ki}\leq\frac{1}{2}\,.\label{partial7}
\end{equation}
In practice, one can consider the full transition matrix at tree-level that connects all possible 2-particle states and impose the bound Eq. (\ref{partial6}) or (\ref{partial7}) on the largest eigenvalue of the matrix, $|\text{Re}\,a^{J}_{fi}|$.

Now the tree-level scattering of the component fields of the fermion multiplets into SM gauge bosons can be classified into coupled channels based on the electric charge conservation. That leads to three charge sectors for $2\rightarrow 2$ channels,
\begin{align}
Q_{\mathrm{tot}}=0:\,\,\, & F_{i}^{(Q)}F_{i}^{(-Q)}\rightarrow W^{+}W^{-},ZZ,\gamma\gamma,\gamma Z,\nonumber\\
Q_{\mathrm{tot}}=1:\,\,\, & F_{i}^{(Q+1)}F_{i}^{(-Q)}\rightarrow W^{+}Z,\, W^{+}\gamma,\nonumber\\
Q_{\mathrm{tot}}=2:\,\,\, & F_{i}^{(Q+2)}F_{i}^{(-Q)}\rightarrow W^{+}W^{+}.\nonumber
\end{align}
In practice, one can determine the largest eigenvalue associated with each coupled-channel transition matrix with $Q_{\mathrm{tot}}=0,1,2$ and impose the bound Eq.(\ref{partial7}) but, as the coupled-channel transition matrix for $Q_{\mathrm{tot}}=0$ contrains more channels into SM gauge bosons compared to the charged sectors, $Q_{\mathrm{tot}}=1,2$, we consider it to determine the possible upper-limit on the $SU(2)$ isospin. Finally, for $Q_{\mathrm{tot}}=0$ sector, the bound given in Eq.(\ref{partial7}) implies that the isospin of the fermionic multiplet in the KNT model (see Appendix \ref{partialapp} for detailed calculation) has to be,
\begin{equation}
j_{F}\leq 8\,\,(\text{one generation})\,\,\,\,\text{and}\,\,\,\, j_{F}\leq 7\,\,(\text{three generations})\,.
\label{partial8}
\end{equation}

One could also include the elastic channels, $F_{i}\overline{F_{i}}\rightarrow F_{i}\overline{F_{i}}$ in the coupled-channel matrix. But the helicity amplitude, with helicity transition $0\rightarrow 0$, in the t-channel via photon exchange has a singularity at $\theta=0$. Such presence of the Coloumb singularity in the elastic channel renders the partial-wave unitarity analysis somewhat ineffective. Therefore, to put more concrete bound on the electroweak quantum numbers, one has to look into the appearance of the low scale Landau pole in the SM gauge couplings in the presence of large electroweak multiplets.

\subsection{Appearence of the Landau pole in gauge coupling}\label{landaupolesec}
The beta-functions of the SM gauge couplings are modified when the larger electroweak fermionic or scalar multiplets are added to its matter content. If the isospin and the hypercharge of the $i$-th fermionic and $j$-th scalar multiplets are $(j_{F_{i}},Y_{F_{i}})$ and $(j_{S_{j}},Y_{S_{j}})$, respectively, the one-loop beta-functions for the $SU(2)_{L}$ and $U(1)_{Y}$ gauge couplings $g$ and $g_{Y}$, respectively become,
\begin{align}
\beta_{g}=\frac{g^{3}}{16\pi^{2}}\left(-\frac{19}{6}+\frac{4}{3}\sum_{i}\kappa_{F_{i}}T(F_{i})+\frac{1}{3}\sum_{j}\eta_{S_{j}}T(S_{j})\right)\,,\label{lan1}\\
\beta_{g_{Y}}=\frac{g_{Y}^{3}}{16\pi^{2}}\left(\frac{41}{6}+\frac{4}{3}\sum_{i}\kappa_{F_{i}}D(F_{i})Y^{2}_{F_{i}}+\frac{1}{3}\sum_{j}\eta_{S_{j}}D(S_{j})Y^{2}_{S_{j}}\right)\,.\label{lan2}
\end{align}
Here, $-\frac{19}{6}$ and $\frac{41}{6}$ in the brackets are the SM contribution to $g$ and $g_{Y}$ beta-functions, respectively. $\kappa_{F_{i}}=1,\frac{1}{2}$ if the $i$-th fermionic multiplet contains Dirac and Weyl fermions, and $\eta_{S_{j}}=1,\frac{1}{2}$ if the $j$-th scalar multiplet contains complex and real scalars, respectively. In addition, the Dynkin index, $T(R)$ of the fermionic or scalar multiplet, $R$ with isospin, $j$ is given by, $T(R)=j(j+1)(2j+1)/3$ and its corresponding dimension as $D(R)=2 j +1$\footnote{The dynkin index, $T(R)$ for representation $R$ is defined, for example in \cite{Gross:1973id, Wilczek:1981iz}, as $C_{2}(R)D(R)=T(R)D(\mathrm{Adj})$ where $D(R)$ and $D(\mathrm{Adj})$ are the dimensions of the representation $R$ and adjoint representation, respectively, and $C_{2}(R)$ is the Casimir invariant associated with representation $R$ given by $C_{2}(R)\mathbb{I}=\sum_{a}T^{a}_{R}T^{a}_{R}$, where $T^{a}_{R}$ are the generators in representation $R$ and $\mathbb{I}$ is an identity matrix of order $D(R)$. For $SU(2)$ representation $R$ with isospin $j$, the dimension is $D(R)=2j+1$ and the Casimir invariant, $C_{2}(R)=j(j+1)$, and the expression for the $T(R)$ follows.}. As the last two terms of Eq.(\ref{lan1}) give a positive contribution to the beta function of $g$, for large enough isospin, they can overcome the negative SM contribution and make it positively large. As a consequence, such large positive beta-function will lead to the appearance of the Landau pole for the $SU(2)_{L}$ gauge coupling at a scale much lower than the Planck scale. As we have not observed non-perturbative weak interaction at energy scale few orders above the electroweak scale, the appearance of such low scale Landau pole put a severe constraint on the size of the electroweak multiplet.

For the KNT model, the additional three fermionic multiplets with $SU(2)$ isospin $j$ and hypercharge, $Y=0$, and one scalar multiplet with isospin $j$ and hypercharge $Y=1$ will contribute to Eq.(\ref{lan1}), and depending on the mass scale, $M_{X}$ where the multiplets are integrated in during the renormalization group running; we tabulate the Landau pole for $g$ in Table \ref{landautab} (see Appendix \ref{landauapp} for details).
\begin{center}
\begin{table}[h!]
\begin{tabular}{|c|c|c|c|c|}
\hline
\backslashbox{{\tiny $SU(2)$ Isospin, $j$}}{{\tiny Mass scale, $M_{X}$}} & $M_{Z}=91.1876$ GeV & $10^{3}$ GeV & $10^{4}$ GeV &  $10^{5}$ GeV\\
\hline
$1$ & $4.6\times 10^{16}$ GeV & $2\times 10^{18}$ GeV & $\gg M_{p}$ & $\gg M_{p}$\\
\hline
$2$ & $9.4\times 10^{3}$ GeV & $1.2\times 10^{5}$ GeV & $1.5\times 10^{6}$ GeV & $1.7\times 10^{7}$ GeV\\
\hline
$3$ & $440$ GeV & $5.1\times 10^{3}$ GeV & $5.4\times 10^{4}$ GeV & $5.8\times 10^{5}$ GeV\\
\hline
\end{tabular}
\caption{Appearance of the Landau pole for $SU(2)_{L}$ gauge coupling with the isospin $j$ in the generalized KNT model. Here, $M_{X}$ is the mass scale where the additional electroweak multiplets are integrated in.}
\label{landautab}
\end{table}
\end{center}
The additional BSM fields of the KNT model which contribute to the neutrino mass generation radiatively at three-loops, can have masses ranging from the electroweak to the TeV scale, and here they are denoted by a common mass $M_{X}$ for simplicity. Therefore, if we consider that the masses of these KNT fields are not too heavy compared to the EW scale, i.e. $M_{X}\simeq M_{Z}$, we can see from Table \ref{landautab} that the Landau pole of the $SU(2)$ coupling $g$ appears at $O(\text{TeV})$ for the 5-plet $(j=2)$ and at 440 GeV for the 7-plet $(j=3)$. In contrast, if masses of the KNT fields are at $O(\mathrm{TeV}-100\,\mathrm{TeV})$ range, i.e. $M_{X}\gg M_{Z}$, and if we consider their contributions to RG running of the $SU(2)$ coupling below $M_{X}$ to be negligible, the Landau poles for 5-plet and 7-plet can be pushed to higher energies by integrating in the KNT multiplets at higher $M_{X}$ but not as high as the Planck scale as we can see for the triplet case $(j=1)$. As the presence of KNT multiplets with $j\geq 1$ eventually leads to the appearance of the Landau pole for $SU(2)$ gauge coupling, $g$ at high energy, we want to know how the KNT mass-scale, $M_{X}$ or integrating in these BSM multiplets at higher energies in our simplified RG running of $g$ would affect the scale of its corresponding Landau pole. But the KNT model's full parameter space contains masses of these additional BSM fields ranging from the EW to multi-TeV scale, therefore from a conservative point of view we can consider the $M_{X}$ close to the EW scale\footnote{In the case of the KNT model, the radiatively generated neutrino mass is directly proportional to the charged lepton masses and the loop function, $\mathcal{F}(M^{2}_{F_{i}}/M^{2}_{\Phi},M^{2}_{S_{1}}/M^{2}_{\Phi})$ which becomes smaller when $M^{2}_{F_{i}}/M^{2}_{\phi}\rightarrow 0$ and $\infty$, and inversely proportional to $M_{\Phi}$. Here $M_{F_{i}}$ and $M_{\Phi}$ are the common masses of the fermion and scalar multiplets (as the mass splittings among the component fields are of $O(100\,\mathrm{MeV})$), respectively,and $M_{S_{1}}$ is the mass of the singly charged scalar singlet. Therefore, for a fixed $M_{\phi}$ around TeV, the fermion mass $M_{F_{i}}\ll M_{Z}$ or $M_{F_{i}}\gg O(\mathrm{TeV})$ would lead to the smaller neutrino masses which would have been incompatible with the experimental bound. So, the low-energy neutrino constraints allow the masses of the KNT fields ranging from close to the EW scale to $O(\mathrm{TeV})$ range. On the other hand, for the neutral component of the lightest fermion multiplet to satisfy the DM relic density constraint via thermal freeze-out mechanism sets the masses of the fermion and scalar multiplets in the TeV range. But one can invoke non-standard DM production mechanisms in the KNT model which can again relax this TeV mass-range requirement. For this reason, we consider the mass scale, $M_{X}$ from close to the EW scale to $O(\mathrm{TeV})$ scales.}. Since we have not seen the signature of new physics in the electroweak scale at energies accessible by the LHC, the appearance of the low-scale Landau pole of the SM gauge coupling for electroweak multiplets with $j\geq 2$ has made the singlet $(j=0)$ and the triplet as the viable BSM addition for the KNT model.

One could improve the analysis of the size of the electroweak multiplet and the Landau pole by using the two-loop beta functions of the gauge couplings as done in \cite{DiLuzio:2015oha, DiLuzio:2018jwd}. However, we refrain from using the two-loop beta functions because firstly, we do not know the UV completion of the KNT model and secondly, the KNT Yukawa coupling $g_{i\alpha}$ enters in the two-loop beta functions of the gauge couplings, and eventually the two-loop Landau pole will also depend on these couplings. For this reason, the size of the electroweak multiplets will not be connected to the appearance of the low scale Landau pole in a straightforward way as in the one-loop case. However, even if we consider the UV completion of the model with large electroweak multiplets in a Grand Unified Theory (GUT) setup, it would require either enormous representations of the minimal GUT like $SO(10)$ or a much bigger GUT group where the model can be embedded in its fundamental or adjoint representations, both of which are not theoretically appealing.

\section{Gamma-ray Flux from Dark Matter Annihilation}\label{DMtriplet}

\subsection{Sommerfeld Enhanced DM Annihilation}\label{SEsec}

In the triplet KNT model, the lightest neutral component of the fermion triplet, $F_{1}^{0}$ is the viable DM candidate. The Sommerfeld enhancement of the DM annihilation to SM gauge bosons takes places when the DM is non-relativistic, $v_{\text{DM}}\ll c$ and the SM gauge bosons follow $m_{W,Z}\ll m_{\text{DM}}$. In this limit, the exchange of W and Z bosons between triplet component fields will lead to Yukawa potentials and $\gamma$ exchange will lead to a Coulomb potential, respectively which in turn significantly modifies the wavefunction of the incoming DM states and enhances the annihilation cross-sections. The calculation of Sommerfeld enhanced DM annihilation cross-section is a well-studied subject, so here we follow the prescriptions given in \cite{Beneke:2014gja}. Nevertheless, we briefly review them to set up our notation.

As the Sommerfeld enhancement is considered for $2\rightarrow 2$ processes, we first define 2-particle states which consist of incoming component fields of fermion triplet. For the DM (co)annihilation channels where the final states consist of $W^{\pm}$, $Z$ and $\gamma$ bosons, the only relevant 2-particle states are CP-even states with total electric charges $Q=0,\,\pm 1,\,\pm 2$. In the case of DM annihilation in the galaxy halo at present times, only 2-particle states with $Q=0$ are applicable.

In general, if the three fermionic triplets of the KNT model are almost mass degenerate, $M_{F_{1}}\sim M_{F_{2}}\sim M_{F_{3}}$, the 2-particle states will contain component fields not only from one multiplet but also from different ones. But, as the hierarchical mass spectrum of $\mathbf{F}_{i}$ is consistent with the best-fit experimental values of the neutrino mixing angles and mass-squared differences \cite{Esteban:2018azc}, it allows us to have $M_{F_{1}}\ll M_{F_{2,3}}$ and $M_{F_{1}}$ to be in O(TeV). It enables us to define 2-particle state vector made out of component fields of $\mathbf{F}_{1}$ as follows,
\begin{align}
Q=0: &\,|\Psi\rangle=(F^{0}_{1}F^{0}_{1},F_{1}^{\pm}F_{1}^{\mp})^{T}\,,
\label{se1}\\
Q=\pm 1: &\,|\Psi\rangle=|F_{1}^{0}F^{\pm}_{1}\rangle\,,\label{se2}\\
Q=\pm 2: &\,|\Psi\rangle=|F^{\pm}_{1}F^{\pm}_{1}\rangle\,.\label{se3}
\end{align}

The modification of the wavefunction which generates the Sommerfeld enhancement, is determined by solving the radial Schrodinger equation with effective potential,
\begin{equation}
\frac{d^2 \Psi_{jj',ii'}}{d r^2}+\left[\left((M_{F_{1}}v)^2-\frac{l(l+1)}{r^{2}}\right)\delta_{jj',kk'}-M_{F_{1}}V_{jj',kk'}\right]\Psi_{kk',ii'}=0\,,
\label{se4}
\end{equation}
where $r$ is the magnitude of the relative distance between two component fields in their center-of-mass frame, the kinetic energy of the incoming DM states, i.e. $|ii'=F^{0}_{1}F_{1}^{0}\rangle$ is $E=M_{F_{1}}v^2$, The wavefunction $\Psi_{jj',ii'}$ gives the transition amplitude from $|ii'\rangle$ states to $|jj'\rangle$ states in the presence of effective potential, $V$. The double indices $ii'$, $jj'$ and $kk'$ run over the states of the 2-particle state vector defined in Eq.(\ref{se1}).

We primarily focus on the S-wave annihilation so we set $l=0$ and have
\begin{equation}
 \frac{d^{2}\Psi_{jj',ii'}}{d r^2}+\left[k^{2}_{jj'}\delta_{jj',kk'}+
 M_{F_{1}}\left(\frac{f_{jj',kk'}\alpha_{a} 
e^{-n_{a}m_{W}r}}{r}+\frac{Q_{kk'}^{2}\alpha_{\text{em}}}{r}\delta_{jj',kk'}
\right)\right]
 \Psi_{kk',ii'}=0\,.
 \label{schrod2} 
\end{equation}
Here, $k^{2}_{jj'}=M_{F_{1}}(M_{F_{1}}v^2-d_{jj'})$ is the momentum associated
with the 2-particle state, $|jj'\rangle$
and $d_{jj'}=m_{j}+m_{j'}-2M_{F_{1}}$ denotes the mass differences between DM and 
other states of
the multiplet. $Q_{kk'}$ is the electric charge associated with state 
$|kk'\rangle$. Also, $\alpha_{W}=\alpha$ and $n_{W}=1$ for W boson exchange and 
$\alpha_{Z}=
\alpha/\cos^{2}\theta_{W}$ and $n_{Z}=1/\cos\theta_{W}$ for Z boson exchange. Finally, $f_{jj',kk'}$ is the group theoretical factor associated with $SU(2)$.

Now by using dimensionless variables defined as 
$x=\alpha m_{F_{1}} r$, $\epsilon_{\phi}=(m_{W}/m_{F_{1}})/\alpha$,
$\epsilon_{v}=(v/c)/\alpha$ and 
$\epsilon_{d_{ii'}}=\sqrt{d_{ii'}/m_{F_{1}}}/\alpha$, we re-write the
coupled radial Schrodinger equations as
\begin{equation}
\frac{d^{2}\Psi_{jj', ii'}}{dx^2}
+\left[\hat{k}_{jj'}^{2}\delta_{jj',kk'}+\frac{f_{jj',kk'}n_{a}^{2}e^{-n_{a}
\epsilon_{\phi}x}}{x}+\frac{Q_{kk'}^{2}\sin^{2}\theta_{W}}{x}\delta_{jj',kk'}
\right]\Psi_{kk',ii'}=0\,,
\label{schrodeq}
\end{equation}
where the dimensionless momentum, 
$\hat{k}^{2}_{jj'}=\epsilon_{v}^2-\epsilon_{d_{jj'}}^{2}$.

At large $x$, $\Psi_{jj',ii'}$ behaves as $\Psi_{jj',ii'}\sim T_{jj',ii'}e^{i\hat{k}_{jj'}x}$ where $T_{jj',ii'}$ is the transition amplitude provided the effective potential is dominated by Yukawa potential. Now if the annihilation matrix for final state $f$ is given by $\Gamma^{(f)}_{jj',ii'}$, the annihilation cross section is,
\begin{equation}
\sigma_{F_{1}^{0}F_{1}^{0}\rightarrow f}=c(T^{\dagger}.\Gamma^{(f)}.T)_{F^{0}_{1}F^{0}_{1},F^{0}_{1}F^{0}_{1}}\,,
\label{se6}
\end{equation}
where $c=2$ for $|F^{0}_{1}F^{0}_{1}\rangle$ state as it consists of identical fields. The Sommerfeld enhancement factor is then given by
$S_{F^{0}_{1}F^{0}_{1}\rightarrow f}=\sigma_{F_{1}^{0}F_{1}^{0}\rightarrow f}/\sigma^{0}_{F_{1}^{0}F_{1}^{0}\rightarrow f}$, where $\sigma^{0}$ is the tree-level annihilation cross-section.

\subsection{Dark Matter Constraints and Parameter Space}\label{dmconstraints}

The relic density of DM in the universe measured by Planck Collaboration as $\Omega_{\text{DM}}h^{2}=0.120\pm 0.001\,(68\%\,\text{C. L.})$ \cite{Aghanim:2018eyx} sets an important constraint for KNT triplet DM. In the case of $F_{1}^{0}$, the standard thermal freeze-out scenario involves the SM gauge interactions and the KNT yukawa interactions given by $g_{i\alpha}$ terms in Eq.(\ref{eq1}).

The DM (co)annihilation into the SM final states, controlled by the SM gauge interactions, take place via $S$-wave and $P$-wave channels. Moreover, the $S$-wave DM (co)annihilation cross-sections receive non-negligible Sommerfeld enhancement. In contrast, the DM annihilation into charged leptons that involves the KNT Yukawa coupling $g_{i\alpha}$, has $S$-wave and $P$-wave channels. However, both of them are suppressed by the light charged lepton mass and the velocity of the DM, respectively. Because of the large multiplicity and unsuppressed Sommerfeld enhanced $S$-wave contribution, the DM (co)annihilation controlled by gauge interaction are more dominant compared to that into charged leptons involving the KNT Yukawa coupling. Therefore, essentially the gauge interaction determines the relic density of the triplet KNT DM in the thermal freeze-out which is calculated using non-relativistic approximation described in \cite{Chowdhury:2018mrr} (and references therein).

\begin{figure}[h!]
\centerline{\includegraphics[width=8cm]{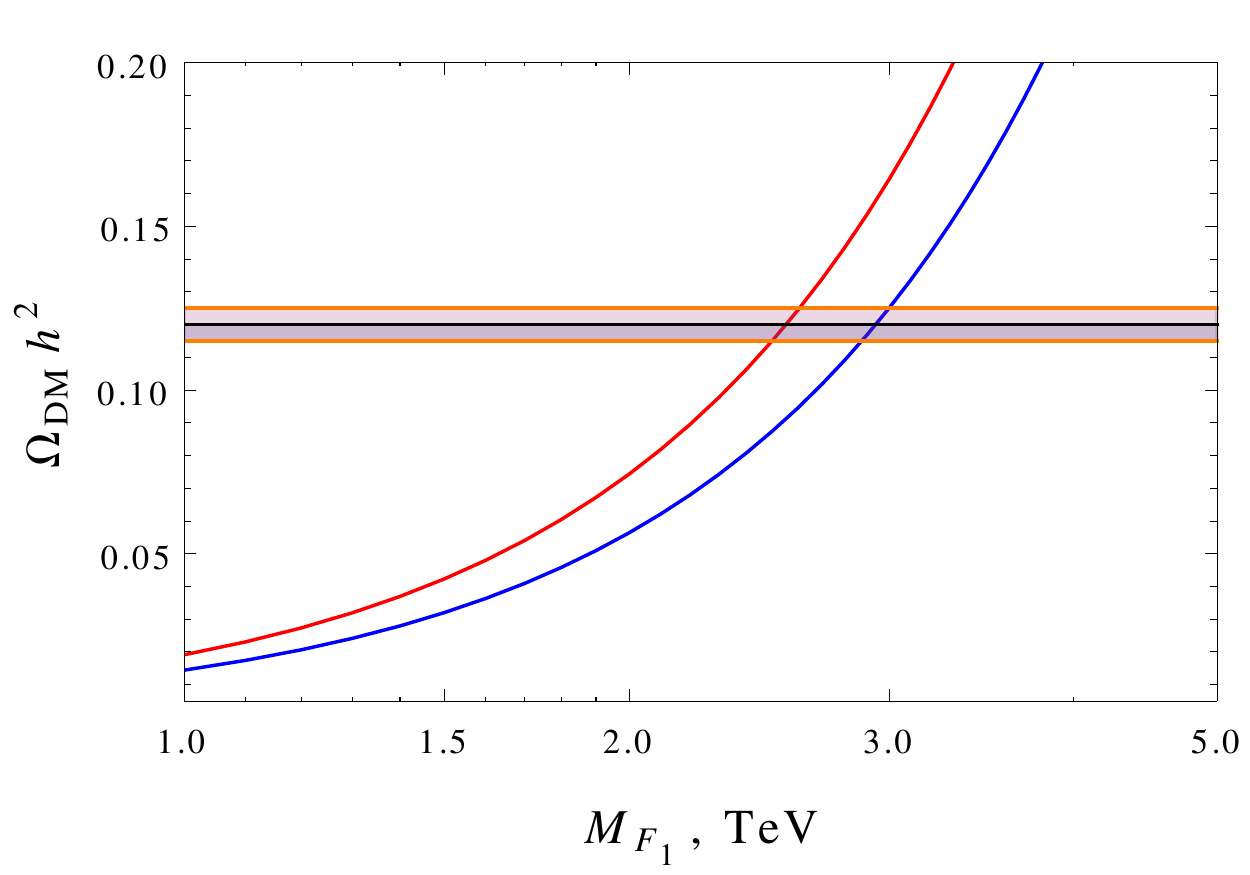}\hspace{0mm}\includegraphics[width=9cm]{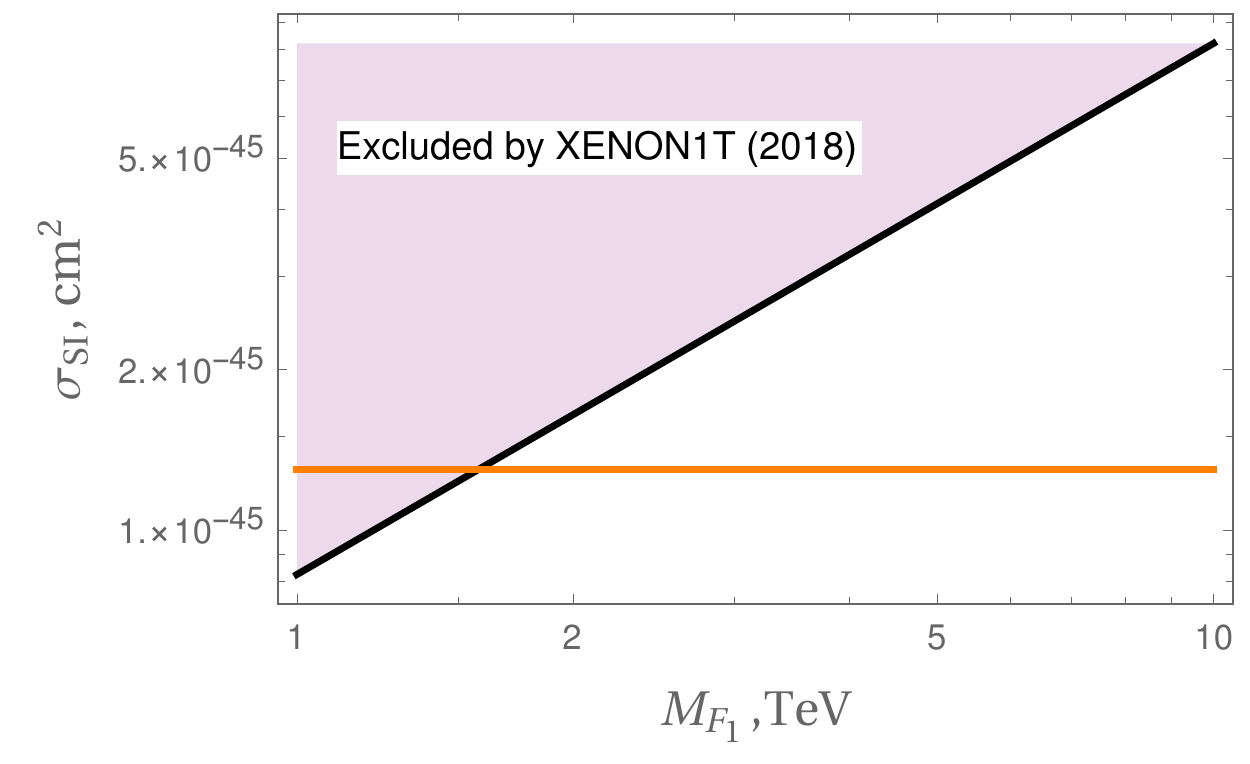}}
\caption{(Left) The DM relic densities, $\Omega h^{2}$ of triplet w/o SE (red) and with SE (blue), respectively. The horizontal band represents the $5\sigma$ band with central value $\Omega h^{2}=0.12$ measured by Planck. (Right) The Spin-independent cross section of triplet DM, $F^{0}_{1}$ and nucleon interaction (orange horizontal line). The shaded region is excluded by XENON1T (2018) data \cite{Aprile:2018dbl}}
\label{dmrelic}
\end{figure}

Another constraint  comes from the DM direct detection. The spin-independent cross-section for the Majarana DM contained in the electroweak multiplet of integer isospin $j$ does not depend on the DM mass, and is given by
\cite{Cirelli:2005uq},
\begin{equation}
\sigma_{\text{SI}}=
j^2(j+1)^{2}\frac{\pi \alpha^2 M_{\text{Nucl}}^{4}f^{2}}{4 m_{W}^{2}}\left(\frac{1}{m_{W}^{2}}+\frac{1}{m_{h}^{2}}\right)\,,
\label{spinind}
\end{equation}
where, $M_{\text{Nucl}}$ is the mass of the target nucleus, $f$ parametrizes nucleon matrix element as $\langle n|\sum_{q}m_{q}\overline{q}q|n\rangle=f\,m_{n}\overline{n}n$ and from lattice result, $f=0.347131$ \cite{Giedt:2009mr}. On the other hand, the spin-dependent cross-section is suppressed by the mass of the DM which is of the $O(\text{TeV})$.

Therefore, we can see that the constraint from the DM relic density sets the mass of the DM in the triplet KNT model as 2.55 TeV (without SE) and 2.94 TeV (with SE), respectively, if the standard thermal freeze-out mechanism is considered. But, as pointed out in \cite{Chowdhury:2018nhd}, the nonthermal decay of the scalar, $\phi^{+}$, can produce the DM $F_{1}$ within the KNT model, although some fine-tuning in the mass-splittings and couplings are needed to get the correct relic density for a wider range of DM masses. However, the fine-tuning can be avoided if the KNT model is coupled to an extended dark sector which would assist the non-thermal production of the DM. For this reason, we probe the DM in the KNT model with mass up to 100 TeV that already includes the mass set by the relic density constraint for the thermal freeze-out scenario, in our CTA sensitivity study.

\subsection{Gamma-ray flux from DM annihilation}\label{gammarayfluxsec}

The gamma-rays from the DM annihilation consist of the prompt gamma-rays and the secondary gamma-rays. The prompt gamma-rays are produced directly from the DM annihilation, or the decays of the SM final states originated in the annihilation. On the other hand, the secondary gamma-rays come from the inverse Compton scattering (ICS) of  $e^{\pm}$ produced in DM annihilation, again directly or from decays of the SM final states, with the ambient photon background, i.e. mainly coming from Cosmic Microwave Background photons, dust rescattered light and starlight. As our analysis focus on the diffuse prompt gamma-ray flux coming from the Sommerfeld enhanced DM annihilation into gauge bosons ($W^{+}W^{-}$) with energy in the TeV range, we neglect the effect of secondary gamma-ray emission from DM annihilation for simplification.

\begin{figure}[h!]
\vspace{-0.425cm}
\centerline{\includegraphics[width=12cm]{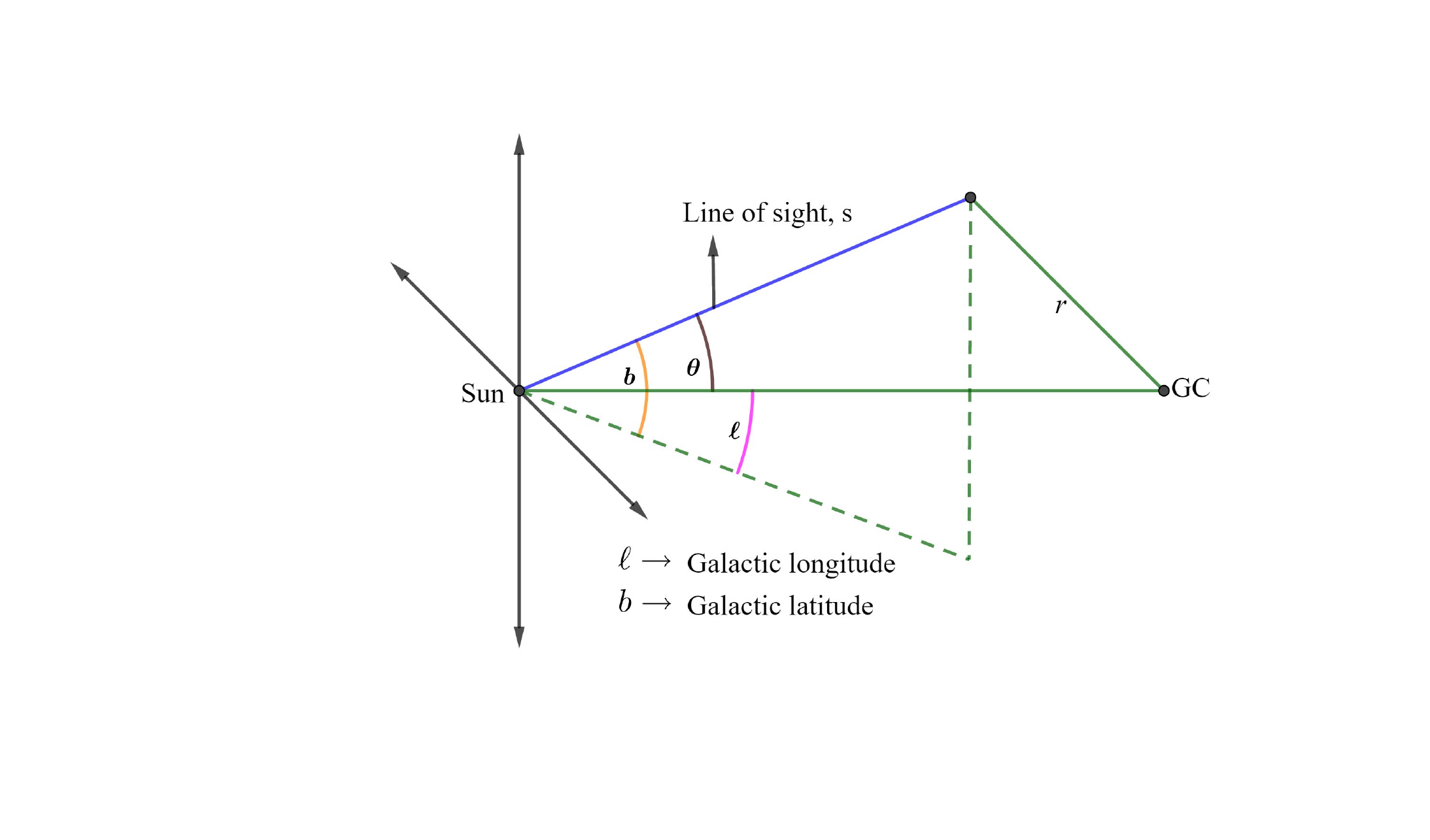}}
\vspace{-2cm}
\caption{The line-of-sight (l.o.s), $s$ for the gamma-ray arrived at the Earth which is originated at a point with distance $r$ from the Galactic center (GC).} 
\label{point}
\end{figure}

The differential prompt gamma-ray flux from the DM annihilation for a given DM mass, $m_{F_{1}}$, in the generalized KNT model is,
\begin{equation}
\frac{d\Phi}{d E_{\gamma}}=\frac{\langle \sigma v\rangle}{8 \pi m_{F_{1}}^{2}}\sum_{f}B_{f}\frac{d N_{f}}{d E_{\gamma}}
\int_{\Delta\Omega}\int_{l.o.s}\rho^{2}_{DM}(r) ds d\Omega\,,
\label{flux1}
\end{equation}
where $\langle \sigma v\rangle$ is the corresponding velocity averaged annihilation cross-section for $m_{F_{1}}$, $d N_{f}/d E_{\gamma}$ is the gamma-ray spectra per annihilation for the annihilation channel, 
$F_{1}^{0} F_{1}^{0}\rightarrow f$ and $B_{f}$ is the corresponding branching ratio. The integration over the solid angle, $\Delta\Omega$ and line-of-sight (l.o.s), $s$ of the squared DM mass density, $\rho_{DM}$ is called the astrophysical J-factor. Here, $r$ is the distance between the Galactic Center (GC) and the point in space characterized by galactic coordinates, $(b,l)$, and it is given as $r(s,\theta)=(r_{\odot}^{2}+s^2-2 r_{\odot} s \cos\theta)^{1/2}$, where $r_{\odot}$ is the distance between the Earth and Galactic Center and $\theta$ is the angle between the line-of-sight and the axis connecting the Earth and the GC as shown in Fig. \ref{point}. In addition, $\theta$ and the galactic coordinates $(b,l)$ are related as $\cos\theta=\cos b \cos l$.

\subsubsection{Gamma-ray Spectrum from DM Annihilation}

\begin{figure}[h!]
\vspace{-0.2cm}
\centerline{\includegraphics[width=10.5cm]{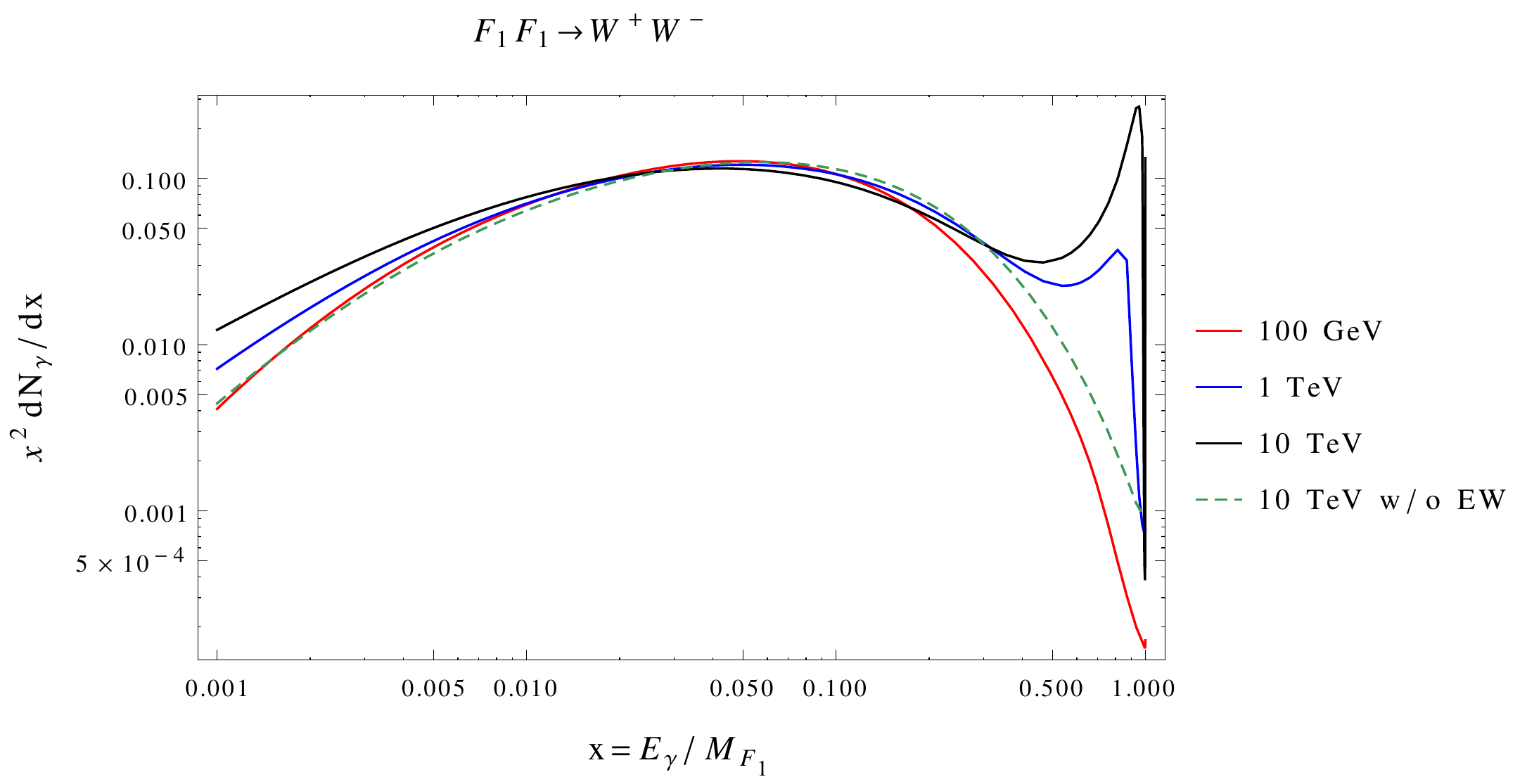}}
\caption{Gamma-ray spectrum per annihilation determined using PPPC4DMID for the DM annihilation channel into $W^{+}W^{-}$.}
\label{dmspec}
\end{figure}

As mentioned already, the dominant DM annihilation channels in the triplet KNT model are those with the SM gauge bosons as final states. Hence, two types of gamma-ray spectra arise from SM gauge boson final states, i) Continuum spectra coming from $W^{+}W^{-}$ and $ZZ$, and ii) line-like spectra from $\gamma\gamma$ and $\gamma Z$ at $E_{\gamma}=M_{F_{1}}$ and $E_{\gamma}=M_{F_{1}}-\frac{M_{Z}^{2}}{4M_{F_{1}}}$, respectively. Besides, the DM annihilation to $\gamma+X$ can also give line-like spectra at the end-point of photon energy. The sensitivity studies for IACTs involving the continuum and line-like photon spectra require different strategies. In our study, we focus on the continuum spectra given by the $W^{+}W^{-}$ final states as a representative channel. We have used PPPC4DMID \cite{Cirelli:2010xx} to calculate the gamma-ray spectra coming from the $W^{+}W^{-}$ final states which include the electroweak (EW) corrections \cite{Ciafaloni:2010ti} that become dominant at $O(\text{TeV})$ energies. From Fig. \ref{dmspec}, we can see the decrease in the gamma-ray spectra per annihilation with the mass of the DM.

\begin{figure}[h!]
\centerline{\includegraphics[width=10cm]{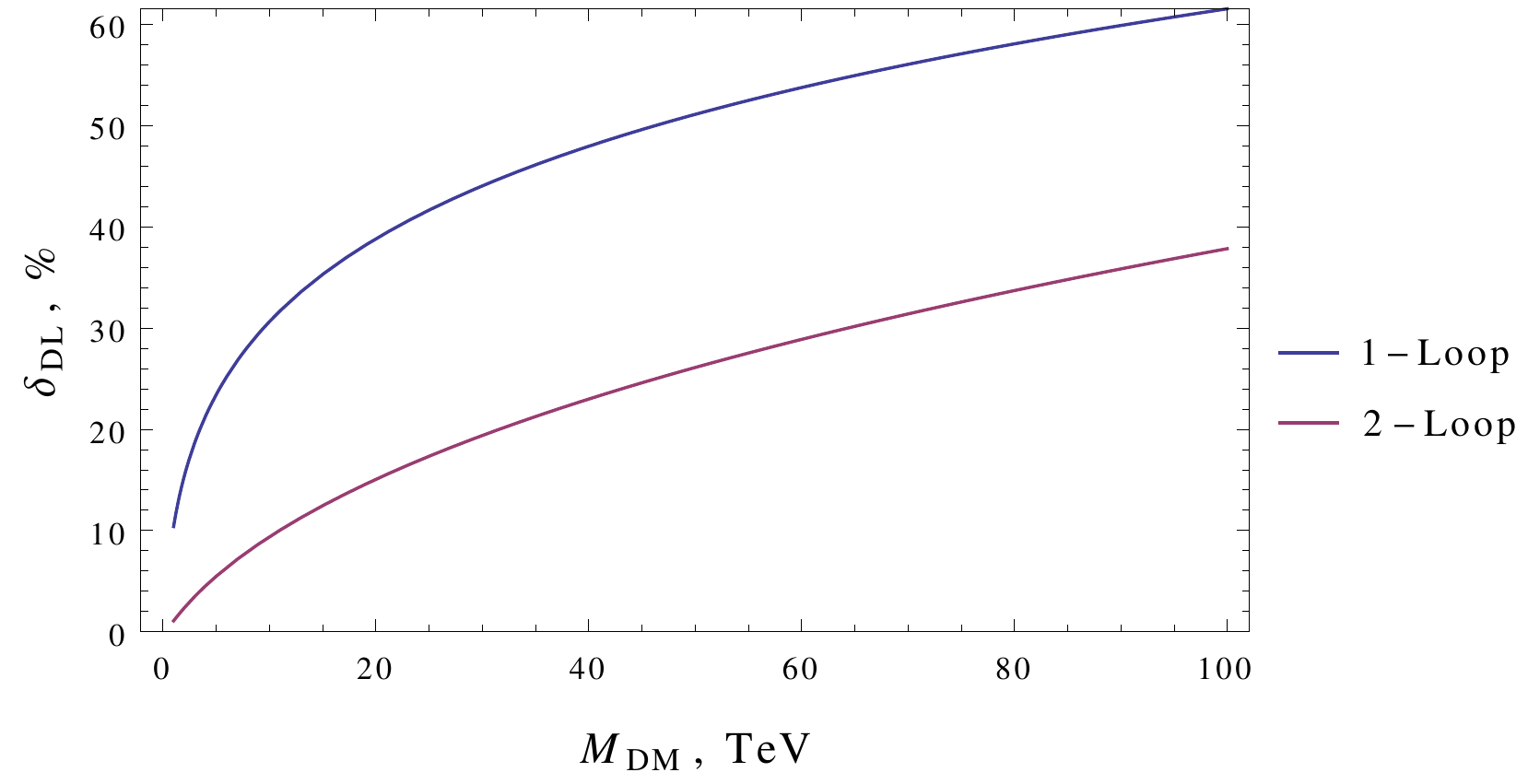}}
\caption{The growing nature of the magnitude of the EW correction, $\delta_{\text{DL}}$ with the DM mass, $M_{DM}$ at one and two loop order necessitates the resummation of EW double logarithms.}
\label{sud}
\end{figure}

In passing, let us make some remarks on the EW corrections implemented in the PPPC4DMID. At energies much higher than the electroweak scale, the dominant EW corrections for a typical $2\rightarrow 2$ process arises from the Sudakov double logarithms which are of the form, $\delta^{(L)}_{\text{DL}}\sim\left(\frac{\alpha}{4\pi}\right)^{L}\text{log}^{2L}\frac{s}{M^{2}_{W}}$ at $L$-th loop order \cite{Denner:2000jv, Denner:2003wi, Manohar:2018kfx}. Here, $\alpha$ is SU(2) coupling and for the non-relativistic DM, the center of mass energy is $s\sim 4M_{\text{DM}}^{2}$. As we can see from Fig. \ref{sud}, the magnitudes of these EW corrections at 1-loop and 2-loop grow with the DM mass, and it indicates that for the DM mass close to or beyond $100$ TeV, one needs to properly resum these Sudakov EW double logarithms and incorporate into the PPPC4DMID. In addition, several studies have been undertaken \cite{Baumgart:2014vma, Ovanesyan:2014fwa, Baumgart:2014saa, Ovanesyan:2016vkk, Baumgart:2017nsr, Baumgart:2018yed, Rinchiuso:2018ajn, Bauer:2020jay} to precisely determine the indirect detection signatures in current and upcoming IACTs for both the line-like and continuum photon spectra coming from TeV-scale or more massive DM annihilation. Besides, the improved estimates of the QCD uncertainties in Monte Carlo event generators might be relevant for the gamma-ray searches of the DM \cite{Amoroso:2018qga}. 

\subsubsection{Astrophysical J-factor}
The astrophysical J-factor plays an essential role in determining the gamma-ray flux from DM annihilation, and therefore, one needs precise information of DM density profile in the Milky Way, especially around the GC which is the focus of this work. The N-body cosmological simulations are often used to parametrize the DM density profile of the Milky Way. The two most frequently used DM density profiles, that assume spherical symmetry, are the Navarro-Frenk-White (NFW) profile and the Einasto profile,
\begin{align}
\rho_{\text{NFW}}(r)&=\rho_{s}\,\frac{r_{s}}{r}\left(1+\frac{r}{r_{s}}\right)^{-2}\,,\label{dmdens1}\\
\rho_{\text{Ein}}(r)&= \rho_{s}\,\text{exp}\left\{-\frac{2}{\alpha}\left[\left(\frac{r}{r_{s}}\right)^{\alpha}-1\right]\right\}\,,\label{dmdens2}
\end{align}
where $\rho_{s}$ and $r_{s}$ are the characteristic density and scale radius, respectively. The NFW profile behaves like $r^{-1}$ at the GC whereas the Einasto profile does not and is smaller in magnitute. The $\alpha$ is the shape parameter that determines the steepness of Einasto profile at the neighbourhood of the GC. The numerical values of these parameters are for NFW profile, $(\rho_{s},r_{s})=(0.184\,\text{GeVcm}^{-3},\,24.42\,\text{kpc})$ and for Einasto profile, $(\rho_{s},r_{s},\alpha)=(0.033\,\text{GeVcm}^{-3},\,28.44\,\text{kpc},0.17)$, respectively.

The value of J-factor is subject to considerable uncertainties because there is no adequate observational data and precise simulation of our galaxy with all its baryonic and DM content. Moreover, the DM density profile depends on the nature of the DM itself (for a review, please see \cite{Salucci:2018hqu}). Therefore, the J-factors used in different studies vary significantly \cite{Pato:2015dua, Hooper:2016ggc, Benito:2019ngh}. Hence, we have used the NFW profile, because of its cuspy nature at GC, as the representative in our CTA  sensitivity study to detect the DM of the KNT model.

\section{Dark Matter Detection with the CTA}\label{ctasec}
\subsection{TeV DM at the CTA}\label{dmcta}

Gamma-rays being electrically neutral, do not deviate from their paths due to the galactic magnetic field when propagating through the galaxy. Moreover, gamma-rays with energies ranging from 100 GeV to 100 TeV, have minimal absorbance in the interstellar medium. So, when these very high energy (VHE) gamma-rays enter the Earth's atmosphere, they undergo collisions with atmospheric molecules and generate an air shower of secondary particles known as Extensive Air Shower. These shower particles descend to Earth with almost the speed of light, and due to the Cherenkov effect, these ultra-relativistic charged particles create a faint blue light in the air which typically lasts for a few nanoseconds. Most of this Cherenkov light is emitted at altitudes ranging between 5 to 15 km, and it propagates down to the ground level as a quasi-planar, thin disk of Cherenkov photons orthogonal to the shower axis. It may cover about $50000\,\mathrm{m}^{2}$ of an area on the ground. By placing arrays of IACTs within the projected Cherenkov light pool, it is possible to detect the air shower provided that the mirror area of the telescope is large enough to catch enough photons. However, it is challenging to reconstruct the exact geometry of the air shower in space with the observation from a single telescope. Hence, multiple telescopes are deployed to take the image of a separate shower from different points which leads a stereoscopic reconstruction of the shower geometry. The images captured by IACTs after removing the backgrounds contributions shows the track of the air shower, which points back to the celestial origin of the incident gamma-rays, and eventually makes it possible to determine the location of its source in the sky along with its spectral and spatial properties.  As we have seen in section \ref{DMtriplet} that when the DM of the triplet KNT model has mass in $O(\text{TeV})$ range, its Sommerfeld enhanced annihilation into SM gauge bosons can produce the VHE gamma-rays of either broad or narrow spectral features. Therefore, the IACTs are ideally suited to detect such TeV DM in the galaxy. 

There are three major currently operational IACTs: H.E.S.S., MAGIC and VERITAS which have performed well within their capabilities and have discovered more than hundreds of VHE gamma-ray sources. However, the decisive point of the scientific performance of CTA will be its ability to survey the sky over broad energy ranges from 20 GeV to 300 TeV with better angular resolution, energy reconstruction and sensitivity, which is one order of magnitude better than existing IACTs. Besides, to provide the full sky coverage, CTA will be installed at two different sites: one in the northern hemisphere at La Palma (Canary Islands, Spain) and another in the southern hemisphere at Paranal (Chile). The southern observatory (known to be as CTA South) will study the GC, and its northern counterpart (known to be as CTA North) will survey extra-galactic objects. As the search for the DM at the GC will be one of the key focus of the CTA \cite{CTAConsortium:2018tzg}, we have chosen to study its detection possibility of the TeV DM of the triplet KNT model and probe its the parameter space.

\subsection{CTA Instrument Response Function}\label{ctairfsec}

The instrument response function (IRF) can be considered as the area times probability that a photon with a given set of input parameters is detected as an event with a set of observables (reconstructed parameters). For CTA, IRF is the product of effective area ($A_{\text{eff}}$), point spread function (PSF), energy resolution and dispersion, and background rate as a function of energy. The key features of these quantities are as follows. 

The effective area is the area within which the CTA can observe air showers. It depends on the energy of the primary gamma-ray and the offset angle $\phi$ (the angle between the array's normal direction and the actual source position). Moreover, the point spread function (PSF) of CTA IRFs represents the spatial probability distribution of reconstructed event positions for a point source. Besides, in short, the energy dispersion of the CTA gives the resolution between the actual and reconstructed gamma-ray energies of events recorded by it. Finally, the background rate is the residual cosmic-ray background rate per solid angle (here, square degree) as a function of reconstructed gamma-ray energy in the CTA.

\begin{figure}[h!]
\centerline{\includegraphics[width=9.15cm]{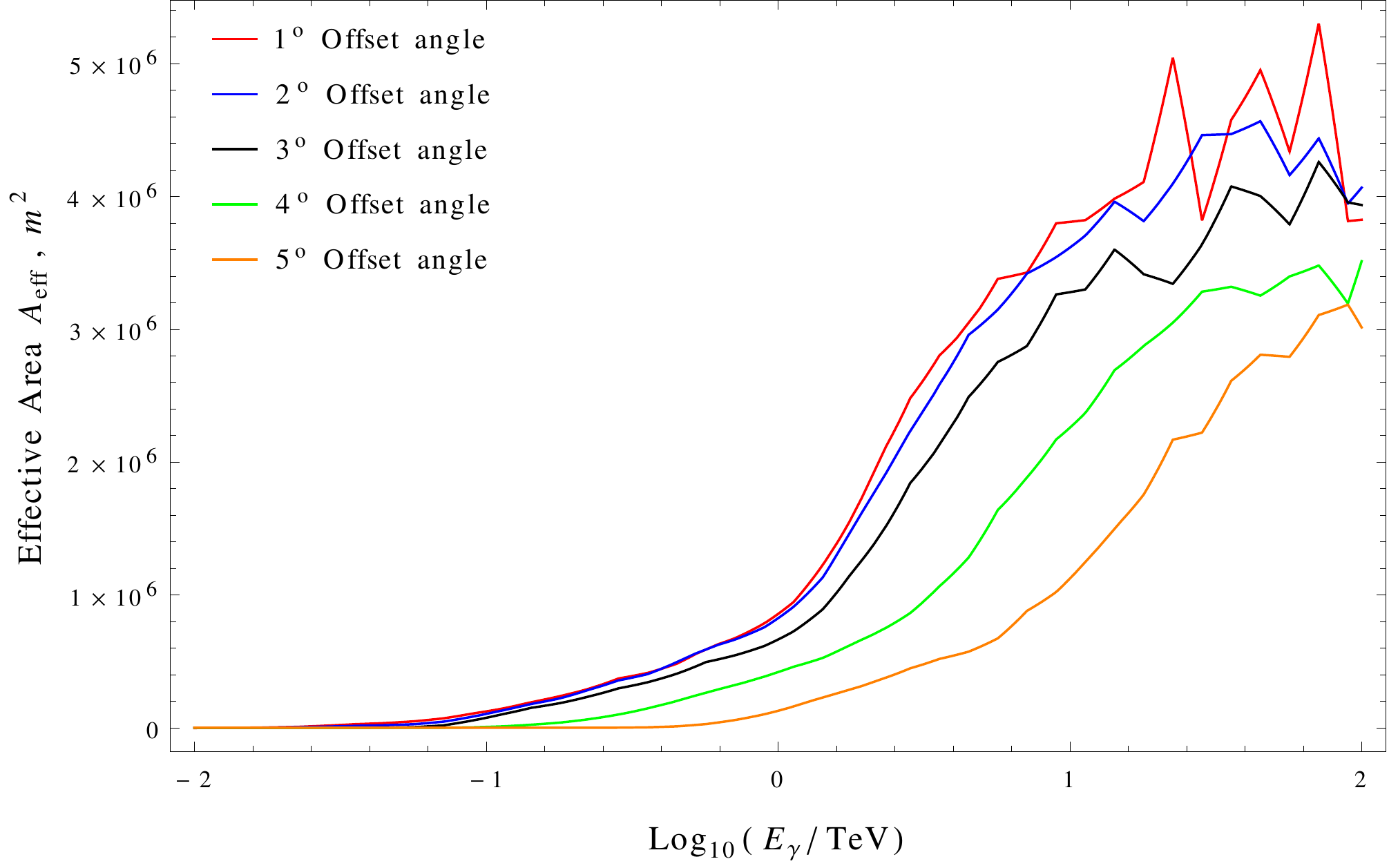}\hspace{3mm}\includegraphics[width=9.15cm]{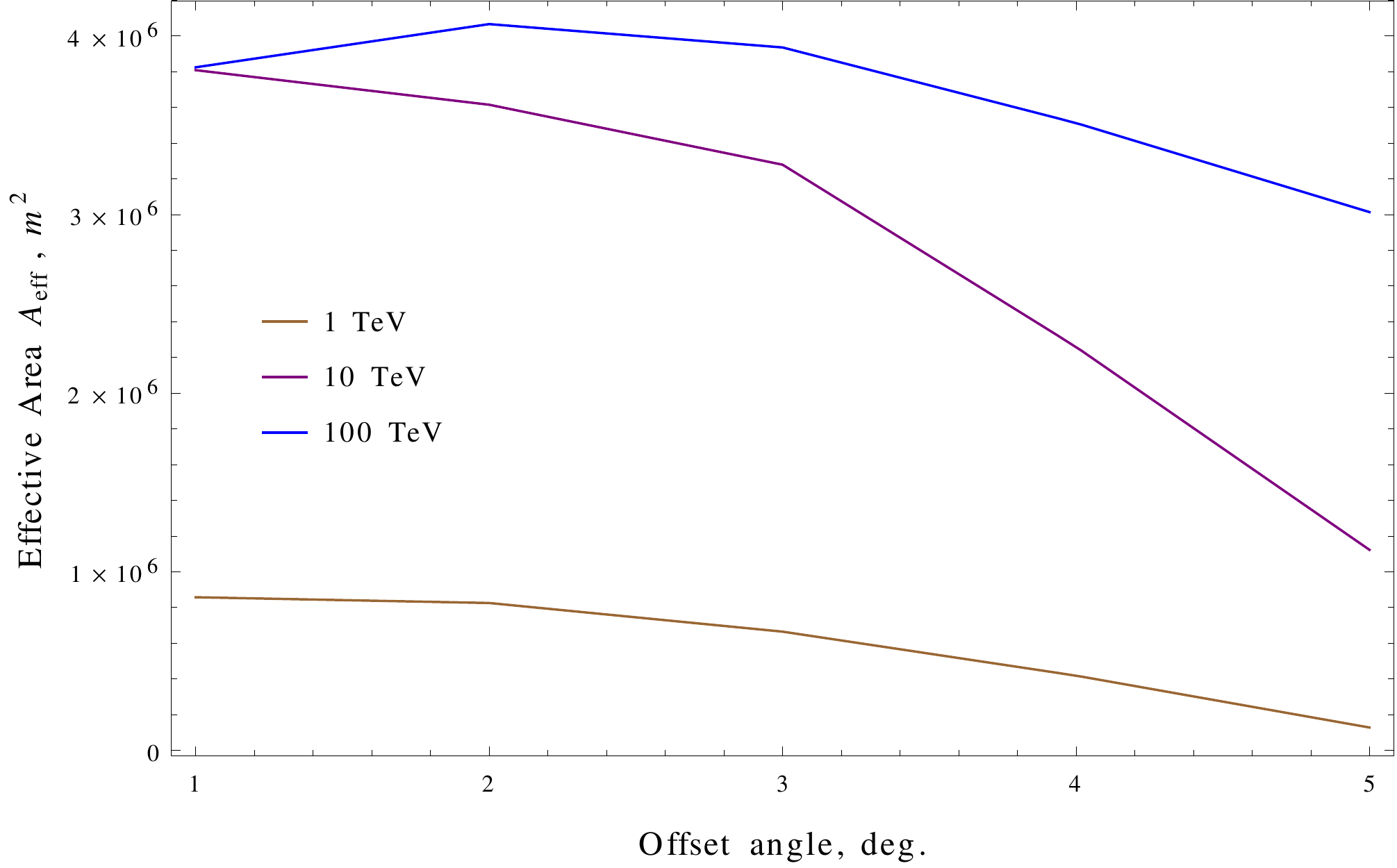}}
\caption{Energy and offset angle dependency of the effective area of IRF for CTA Southern site with zenith angle $20^{\circ}$ and exposure time 50 hours (\texttt{S\_z20\_50h})}
\label{dmeffarea}
\end{figure}

As the CTA is not constructed yet, the CTA consortium has carried out several Monte Carlo simulations \cite{Bernlohr_2013, Hassan_2017, Acharyya_2019} to compute them. We specifically make use of $\mathbf{South\_z20\_50h}$ in \textbf{CTA-Performance-prod3b-v2-FITS.tar.gz} IRF (southern site with zenith angle $20^{\circ}$ and exposure time 50 hours) throughout our entire study. Let us point out a few factors for choosing this IRF. Due to its privileged location, the CTA southern site is most favorable for surveying the $\gamma$ sources in GC with energy ranges from 20 GeV to 300 TeV. Since our study is focused on the DM signal from the GC, we considered the IRFs designed for the southern site. Since it is assumed that the total operation time of CTA per year will be 1000 hours, we, therefore, expect that a longer exposure period is necessary to probe a possible a DM source region.
From Fig. \ref{dmeffarea}, we can see the variation of the effective area, that will enter into our computation of gamma-ray counts due to DM or background in the CTA, with respect to gamma-ray energy and off-set angle.

\subsection{CTA Backgrounds}\label{ctabackgroundsec}
In this section, we summarize the dominant background sources at the CTA considered for our study of detecting the DM signal from the KNT model. 

\begin{enumerate}
  \item \textbf{Cosmic Rays}: The Cosmic Rays (CRs) are highly energetic protons and nuclei. Apart from the VHE gamma-rays, CR also trigger air showers, which are $10^3$ times larger than that of $\gamma$ rays, and will act as the dominant background for the CTA. The CR, during their flight from their origin, undergo deflection due to galactic magnetic fields; only particles with sufficiently high energies can reach the Earth's atmosphere. While entering the Earth's atmosphere, CR protons undergo inelastic collisions with atmospheric air molecules, causing mixed hadronic and electromagnetic air showers \footnote{There is a certain minimum threshold energy required in order to trigger the Cherenkov light shower by VHE-$\gamma$ and CR. For electrons, muons, pions and protons the threshold energies are 21 MeV, 3.4 GeV, 5.6 GeV and 38 GeV, respectively.}. However, because of the large transverse momentum transfer due to hadronic interactions, the shower components are broad and exhibit irregular patterns compared to those of electromagnetic showers initiated by the primary gamma-ray.

The shape of the captured image induced by the Cherenkov light from the air shower initiated by VHE gamma-ray can be well approximated by an ellipse. But the image gathered from the CR induced air shower has a distorted elliptical shape which can allow one to discriminate events generated due to gamma-rays and cosmic rays. Nevertheless, a small fraction of images remain indistinguishable from the signal gamma-rays, which constitute the irreducible background for the CTA. 

 \item \textbf{Galactic Diffuse Emission}: Other than CR, Galactic Diffuse Emission (GDE) acts as a potential background for CTA, which originates from the interaction of CR with interstellar molecules and atomic gas filling the Galactic plane. Observation of \texttt{Fermi-LAT} found that up to energy scales of 100 GeV, the GDE is dominated by $\pi^{0}$ decay, inverse Compton scattering and Bremsstrahlung radiation \cite{Ackermann_2012}. The contribution due to GDE becomes significant when the region of interest becomes close to the GC, which is relevant for our case since we are studying the possible DM sources near the GC.   
  
Modelling TeV range GDE for existing ground-based IACTs is a challenging task due to the uncertainties associated with the estimated background in the telescopes' field of view. It is also impossible to find a signal-free region in the sky by apriori consideration \cite{Neronov:2019ncc}. Besides, the data available from current experiments is limited to have a proper modelling of GDE at the 100 TeV energy scale. As a consequence, we have taken a simplied approch following \cite{Silverwood:2014yza, Lefranc:2015pza, Balazs:2017hxh} and left more realistic GDE modelling for a future work. We incorporate the GDE in our analysis using the P7V6 model of Fermi team \footnote{https://fermi.gsfc.nasa.gov/ssc/data/access/lat/BackgroundModels.html} which fits extremely well for 50 MeV and 500 GeV energy range, and use a power-law extrapolation of P7V6 data to 100 TeV. The H.E.S.S. observation of GDE from Galactic ridge for Galactic coordinates $|b|<0.3^{\circ}$ and $|l|< 0.8^{\circ}$ \cite{Aharonian:2006au, Abramowski:2014vox} is not taken into account as it falls within the region that we have excluded from our study. 
\end{enumerate}

\subsection{Region of Interest}\label{obsregionsec}

\begin{figure}[h!]
\centerline{\includegraphics[width=6cm]{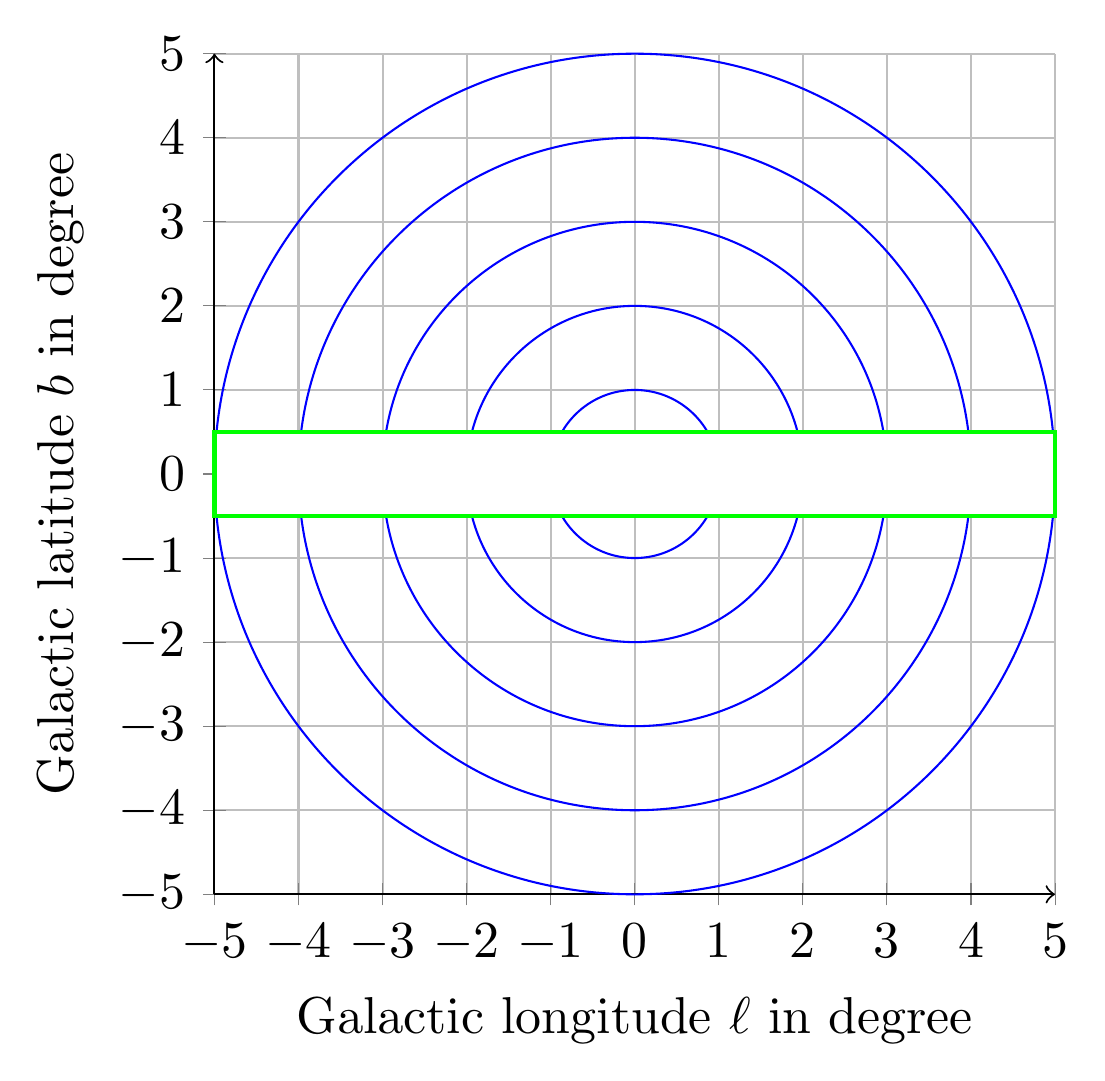}\hspace{0mm}\includegraphics[width=9cm]{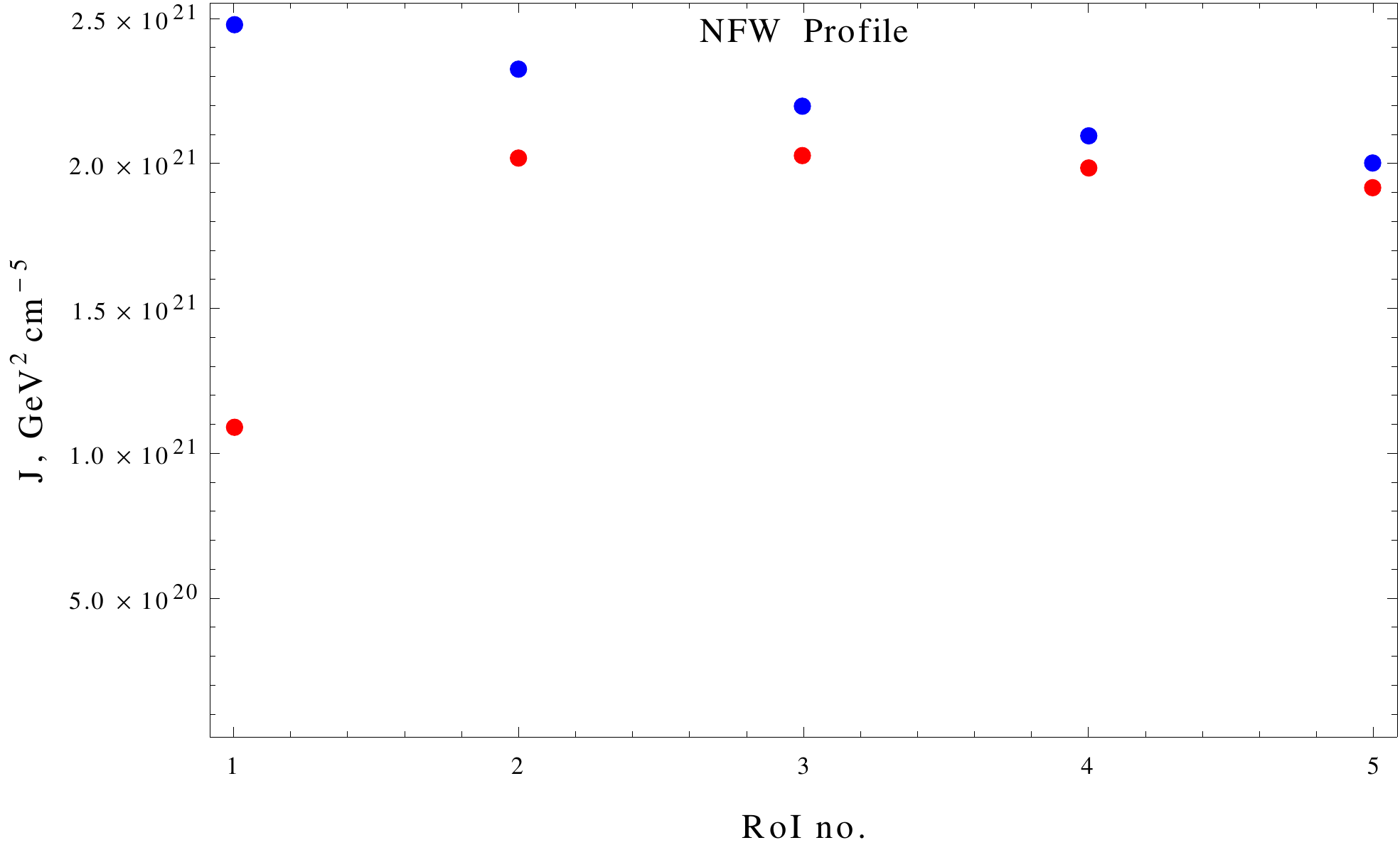}}
\caption{(Left) The annular region of interest (RoI), each with $1^{\circ}$ width, and where the rectangular region $0.3^{\circ} < b < -0.3^{\circ}$ and $-5^{\circ} < \ell< 5^{\circ}$ is excluded from each annular ring because of large astrophysical background. (Right) J-factor corresponding to each annular region of interest, RoI, determined for the NFW DM profile. Here, the blue points indicate the J-factor values for each annular ring. On the other hand, the red points are the J-factor values after subtracting the contribution from the central rectangular patch.
}
\label{jfactor}
\end{figure}

Our observation method is based on the Multi-RoI morphological analysis \cite{Lefranc:2015pza, Balazs:2017hxh}. We assumed that all telescopes are pointing toward the GC with Galactic longitude $\ell$ and latitude $b$ with coordinates $(0,0)$. The Region of Interest (RoI) is divided into five concentric circles, each with width $1^{\circ}$ such that the outermost circle has radius $5^{\circ}$, as seen in Fig. The expected number of photon counts for \textbf{signals} (DM) and \textbf{backgrounds} (CR, GDE) in each circular RoI has been computed simultaneously. It is known that the central part of the GC is populated with several astrophysical sources including Sagittarius A* (SgrA*), emitting gamma-rays. In order to exclude photon counts coming from this region we disregard the central part of the GC by introducing a rectangular patch with $0.3^{\circ} < b < -0.3^{\circ}$ and $-5^{\circ} < \ell< 5^{\circ}$ within the RoI. The corresponding value of the J-factor, considering the NFW profile, is shown in Fig. \ref{jfactor}.

\section{Analysis and Results}\label{resultssec}

\subsection{Expected Gamma-ray Counts at the CTA}\label{expcountsec}
In this section, we obtain the expected counts for the dark matter, cosmic rays and galactic diffuse emission using the methods described below.

The expected differential count for each RoI \textit{i} and energy bin \textit{j} for the source $X$ is calculated using the relation,
\begin{equation}
\frac{d \Gamma_{\gamma,i}^{X}}{d E'_{\gamma}}=
\int_{\Omega'_{i}}d\hat{p}'\int d E_{\gamma}\int d\hat{p}\,\text{A}_{\text{eff}}(E_{\gamma},\hat{p})\,PSF(\hat{p},\hat{p}')E_{\text{disp}}(E'_{\gamma},E_{\gamma},\hat{p})\frac{d\phi_{\gamma}^{X}}{d E_{\gamma}d\Omega}(E_{\gamma},\hat{p})\,,
\label{diffcount}
\end{equation}
where the differential gamma-ray flux for the component $X$, $\frac{d\phi_{\gamma}^{X}}{d E_{\gamma}d\Omega}$ is weighted by the effective area $\text{A}_{\text{eff}}(E_{\gamma},\hat{p})$, point spead function, $PSF(\hat{p},\hat{p}')$ and energy dispersion, $E_{\text{disp}}(E'_{\gamma},E_{\gamma},\hat{p})$. Here, $(E_{\gamma},\hat{p})$ and $(E'_{\gamma},\hat{p}')$ denote the energy and direction of the actual and reconstructed gamma-rays, respectively. The $PSF(\hat{p},\hat{p}')$ which is a probability distribution function of the angular separation between the actual and reconstructed gamma-rays, is mostly relevant for a point source in the sky but the sources we are considering, i.e. DM, CR or GDE, are extended so, the PSF can be well approximated by a delta-function. Moreover, the energy dispersion, $E_{\text{disp}}$ is important if we are looking for a specific energy of the gamma-ray, which applys to the line-like spectrum from DM annihilation. But, in our case, the resulting gamma-ray spectrum from DM annihilation has continuum spectrum so again weighting the differential gamma-ray flux with gaussian-like energy dispersion function is not important. Finally the  photon count is obtained using
 \begin{equation}
     \mu_{ij}^{X} = T_{\mathrm{obs}}\int_{\Delta E_{j}}\mathrm{d}E_{\gamma}\frac{\mathrm{d}\Gamma^{X}_{\gamma,i}}{\mathrm{d}E_{\gamma}}\,,
\label{count}
 \end{equation}
where, $T_{\textrm{obs}}$ is the total observation time, and the integration region $\Delta E_j$ denotes the $j$-th energy bin.

In our analysis, we divide our photon counts into twenty logarithmically spaced energy bins corresponding to the photon energy range, $E_{\gamma}=30$ GeV - $M_{F_{1}}$ where the DM mass, $M_{F_{1}}$  again varies from 1 TeV to 100 TeV, and five RoIs as described in section \ref{obsregionsec}. Also, the observation time is set to $T_{\text{obs}}=100$ hr. For the DM annihilating into $W^{+}W^{-}$ final states, the associated photon count is computed using Eq.(\ref{flux1}), (\ref{diffcount}) and (\ref{count}) for the NFW profile .

Again, we compute the expected photon count for the GDE using Eq.(\ref{diffcount}) and (\ref{count}) from our simplified gamma-ray flux for the GDE which is a power-law extrapolation up to 100 TeV from the Fermi P7V6 model,
\begin{equation}
\frac{\mathrm{d}\phi^{\mathrm{GDE}}}{\mathrm{d}E_{\gamma}}=1.0064\times 10^{-6}\left(\frac{E_{\gamma}}{\text{GeV}}\right)^{-2.333}\,\,\text{GeV}^{-1}\text{cm}^{-2}\text{s}^{-1}\text{sr}^{-1}\,.
\label{gdeeq}
\end{equation}

\begin{figure}[h!]
    \centering
\includegraphics[width=18cm]{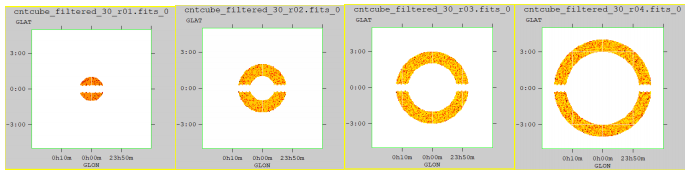}
    \caption{Expected gamma-ray counts due to the CR background determined by the \texttt{ctools} from the first four RoIs where the energy range is from 30 GeV to 30 TeV.}
    \label{CTOOLSfig}
\end{figure}

The expected gamma-ray count associated with the CR is evaluated using \texttt{ctools} version 1.6.2 \cite{Knodlseder:2016nnv} which is a software toolkit designed for data analysis of the CTA and other IACTs. The \texttt{ctools} consists of a set of binary executable C++ and python tools for performing the necessary data analysis where each of these tools needs a set a parameter like telescope pointing directions, radius of FoV, pixelation, time period of observation, calibration datebase etc. Equipped with IRF-\texttt{South\_Z20\_50h}, we use four components of the \texttt{ctools} in our analysis: \texttt{ctobssim}, \texttt{ctbin}, \texttt{ctcubemask} and \texttt{csresspec} following the sequence,
\begin{center}
\texttt{ctobssim} $\rightarrow$ \texttt{ctbin} $\rightarrow$ \texttt{ctcubemask} $\rightarrow$ \texttt{csresspec}.\\
\end{center}
Although \texttt{csresspec} inspects the spectral fit residual in \texttt{ctools}, we use it to extract the count per energy bin. Besides, \texttt{ctcubemask} of the \texttt{ctools}-1.6.2 does not have an option to carry out rectangular masking of the region: $0.3^{\circ} < b < -0.3^{\circ}$ and $-5^{\circ} < \ell< 5^{\circ}$. To do this task, we exclude the region from longitude $-5^{\circ}$ to $5^{\circ}$ using overlapping circles of radius $0.3^{\circ}$. The output from \texttt{ctools} of the expected gamma-ray counts due to the cosmic-rays from first four RoIs are shown in Fig. \ref{CTOOLSfig}.

\begin{figure}[h!]
\centerline{\includegraphics[width=12cm]{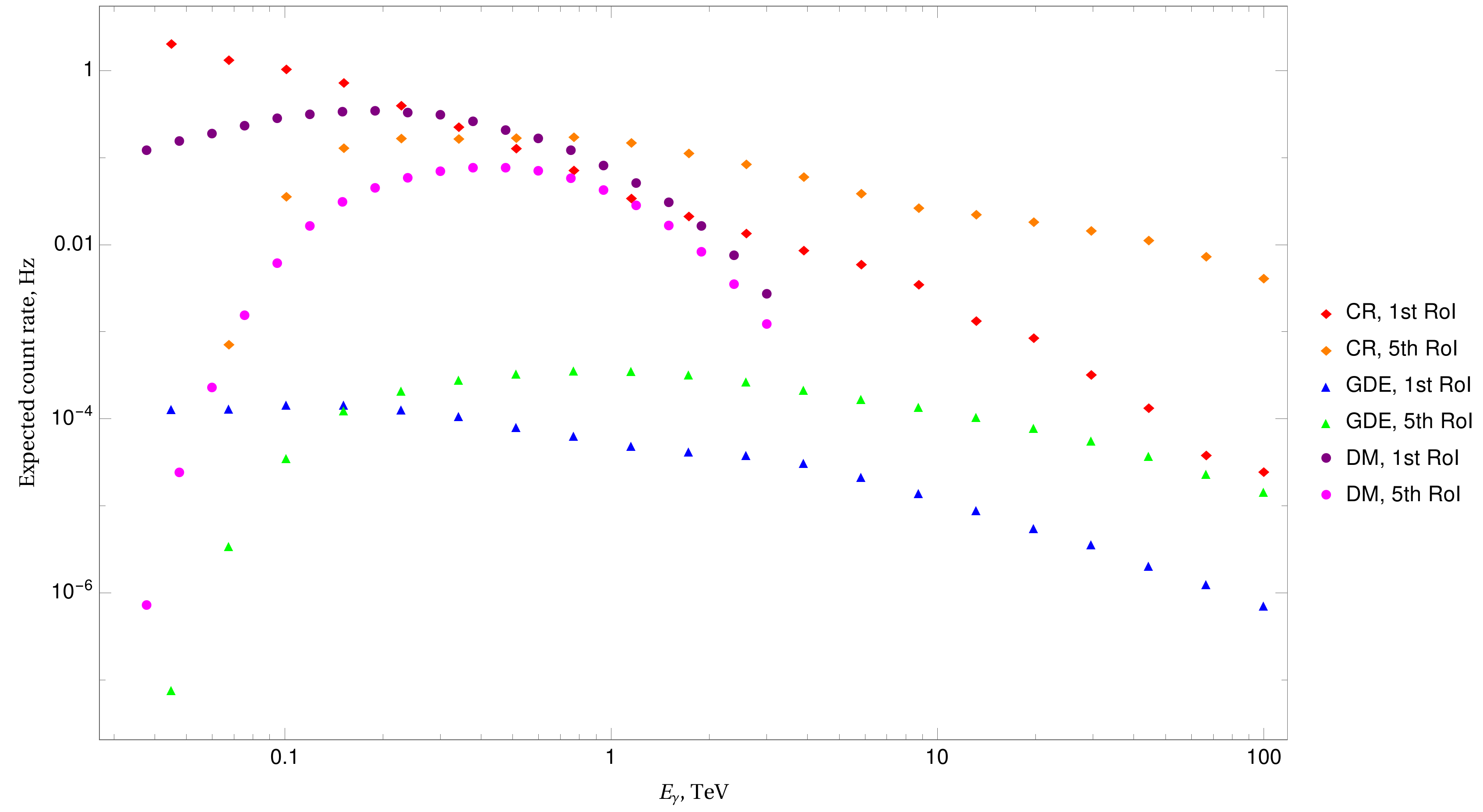}}
\caption{Expected count rate from the DM, CR and GDE with respect to the gamma-ray energy, $E_{\gamma}$ at the CTA for the 1st and 5th RoIs, respectively. Here, the SE annihilation cross-section $\langle\sigma v\rangle=2.6\times 10^{-23}\text{cm}^{3}\text{s}^{-1}$ for DM mass, $M_{F_{1}}=3$ TeV is used to calculate the expected count rate for the DM.}
\label{countvsen}
\end{figure}

In addition, as seen in Fig. \ref{countvsen}, for both 1st (innermost) and 5th (outermost) RoI, the Sommerfeld enhancement in the DM annihilation allows its count rate to be comparable with the CR count. Moreover, we can see that in the case of CR and GDE, at low energy, the expected count rates are smaller in the 5th RoI than that in the 1st RoI. On the contrary, when the gamma-ray energy becomes higher, the count rate in the 5th RoI becomes larger compared to the 1st RoI. In Fig. \ref{dmeffarea} one can see that for low energy, the effective area of the CTA is small and increases with the gamma-energy. Besides, the incident gamma-ray at a large angle with respect to the CTA telescope axis needs to have comparatively higher energy to initiate Cherenkov light that is detectable at the CTA. For this reason, we see a low count rate at low energy in the 5th RoI. On the other hand, with increasing angle towards the outer RoIs, the combination $\Omega_{i}\times A_{\text{eff}}$ becomes large at higher energies, and therefore shows higher gamma-ray counts at high energies for the CR and GDE. In the case of DM, the expected count rate is proportional to the J-factor, which slightly decreases towards outer RoI, as seen in Fig. \ref{jfactor}. For that reason, we see that the expected count rate for DM is large to some extent in the 1st RoI compared to the 5th one.

\begin{figure}[h!]
\centerline{\includegraphics[width=10cm]{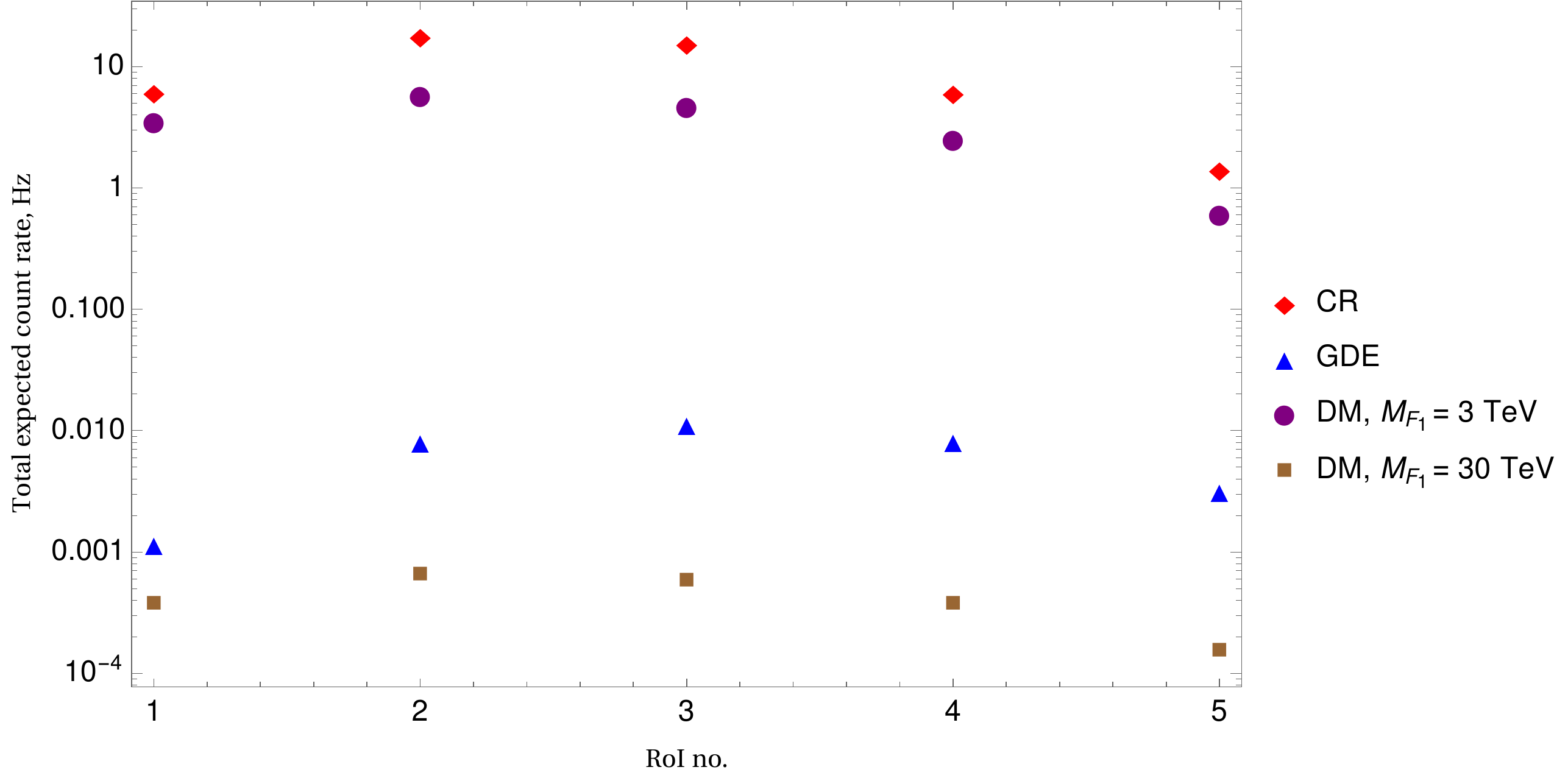}}
    \caption{Total expected count rates from DM, CR and GDE at each RoI. Here the total expected count rate at $i$-th RoI is given by $\frac{\sum_{j}\mu_{ij}}{T_\text{obs}}$. Moreover, as the SE annihilation cross-sections for DM masses 3 TeV and 30 TeV are $\sigma v=2.6\times 10^{-23}\text{cm}^{3}\text{s}^{-1}$ and $\sigma v=1.3\times 10^{-26}\text{cm}^{3}\text{s}^{-1}$, respectively, the total expected count rates for 3 TeV DM are comparable to CR counts.}
    \label{totalexpcountrate}
\end{figure}

The impact of the Sommerfeld enhancement on the DM annihilation is seen again in Fig. \ref{totalexpcountrate} where we consider the total expected count rate in each RoI for the DM, CR and GDE. In this case, the variation of the expected count rate with respect to gamma-ray energy at each RoI is averaged out.
 
\subsection{Likelihood Analysis}\label{likelihoodsec}

In this section, we combine the expected gamma-ray counts coming from DM, CR and GDE backgrounds at the CTA, and determine its prospect to detect the DM in the triplet KNT model.

We divide our photon counts into twenty logarithmically spaced energy bins and five RoIs in a procedure similar to \cite{Silverwood:2014yza, Balazs:2017hxh} using the binned Poisson likelihood analysis. The counts are labeled with $\mu^{X}_{ij}$ where $X$ indicates the source - DM, CR, or GDE, and the lower indices indicate the count in the $i$-th energy bin and $j$-th RoI. The number of expected counts is then given by

\begin{align}
\mu_{ij}&=\mu^{\mathrm{DM}}_{ij}+R^{\mathrm{CR}}_{i}\mu^{\mathrm{CR}}_{ij}+R^{\mathrm{GDE}}_{i}\mu^{\mathrm{GDE}}_{ij}\,,\label{TotalCount}
\end{align}
where, we have introduced the rescaling parameters $\{\mathbf{R}^{\mathrm{CR}} , \mathbf{R}^{\mathrm{GDE}}\}$. We may now write the likelihood function as the product of the independent Poisson distributions associated with each energy bin and RoI:
\begin{align}
L(\bm{\mu}, \bm{R}^\mathrm{CR}, \bm{R}^\mathrm{GDE}| \bm{n})=\prod_{ij} \frac{\mu_{ij}^{\ n_{ij}}}{n_{ij}!}e^{-\mu_{ij}}\,,\label{LikelihoodFunction1}
\end{align}
where, $n_{ij}$ represents the number of observed counts. However, we do not simulate individual Poisson realizations of the observed counts $n_{ij}$, but instead obtain an \textit{Asimov data set} \cite{Cowan:2010js}. In this procedure, the entire observed count is considered to be the sum of CR and GDE backgrounds only, and it can be found by setting $\mu^{\mathrm{DM}}_{ij}\rightarrow 0$ and ${R^{\mathrm{CR/GDE}}_{i}\rightarrow 1}$ in Eq.\eqref{TotalCount}. 

Furthermore, we account for the systematic uncertainties in the signals by introducing a new set of Gaussian distributed nuisance parameters $\alpha_{ij}$, with mean 1 and standard deviation $\sigma_{\alpha}$. This approach results in a modification of Eq.\eqref{LikelihoodFunction1} so that the new likelihood function is
\begin{align}
L(\bm{\mu},\bm{\theta} | \bm{n})=\prod_{ij} \frac{(\alpha_{ij} \mu_{ij})^{\ n_{ij}}}{\sqrt{2\pi} \sigma_{\alpha} n_{ij}!}e^{-\alpha_{ij} \mu_{ij}}e^{-\frac{(\alpha_{ij}-1)^2}{2\sigma_{\alpha}^2}}\,,\label{LikelihoodFunction2}
\end{align}
containing the nuisance parameters we denote collectively by the set $\bm{\theta}=\{\bm{\alpha},\bm{R}^\mathrm{CR},\bm{R}^\mathrm{GDE}\}$.

We separate our analysis into three distinct parts:
\begin{enumerate} 
\item
first, we consider neglecting systematic uncertainties completely, as well as neglecting the photon count from GDE ($\bm{\alpha}\rightarrow 1, \sigma_{\alpha}\rightarrow 0, \mu^\mathrm{GDE}_{ij}\rightarrow 0$); 
\item
next, we consider the effect of including GDE while still neglecting systematic uncertainties ($\bm{\alpha}\rightarrow 1, \sigma_{\alpha}\rightarrow 0$); 
\item
finally, we consider systematic uncertainties restricted to $1\%$ ($\sigma_{\alpha}\rightarrow 0.01$). 
\end{enumerate}
We now use maximum likelihood estimates of the nuisance parameters separately in each of the above three cases to evaluate the test statistic
\begin{align}
\lambda (m_{F_{1}},\langle \sigma v \rangle )=-2\log\Bigg(\frac{L(\bm{\mu},\bm{\theta'}_{\text{MLE}}|\bm{n})}{L(\bm{\mu}_{\text{MLE}},\bm{\theta}_{\text{MLE}}|\bm{n})}\Bigg)\,,\label{TS}
\end{align}
where the subtext MLE indicates maximum likelihood estimates of the underlying parameters and we suppress the implicit dependence on the DM mass, $m_{F_{1}}$ and cross-section, $\langle \sigma v \rangle$ in $\bm{\mu}$ for notational clarity (Recall that $\mu^\mathrm{DM}_{ij}\sim \langle \sigma v \rangle$). $\lambda$ is asymptotically equivalent to a $\chi^2$ distribution with one degree of freedom (since $\bm{\mu}$ in Eq.\eqref{TS} is still a free parameter). We bound the rescaling factors in Eq.\eqref{TotalCount} so that $0.5 \leq R^\mathrm{CR}_{i} \leq 1.5$ and $0.2 \leq R^\mathrm{GDE}_{i} \leq 5$ (in accordance with \cite{Silverwood:2014yza}) and we require $\bm{\alpha} \geq 0$ in order to keep probabilities non-negative in Eq.\eqref{LikelihoodFunction2}. While we formally maximize the likelihood function, in practice we minimize the negative log-likelihood function:
\begin{align}
-\log L (\bm{\mu},\bm{\theta}|\bm{n})= -\sum_{ij}\Bigg( n_{ij}\log (\alpha_{ij}\mu_{ij})-\alpha_{ij}\mu_{ij}-\frac{(\alpha_{ij}-1)^2}{2\sigma_{\alpha}^2}\Bigg)\,,
\label{NegativeLogLikelihood}
\end{align}
where we neglect constants and terms involving $n_{ij}$ which ultimately cancel in evaluating $\lambda$ and hence do not affect our analysis. Furthermore, we reduce the set of nuisance parameters by first analytically maximizing $\log L$ with respect to $\alpha_{ij}$:
\begin{align}
\frac{\partial \log L}{\partial \alpha_{ij}}(\bm{\mu},\bm{\theta}|\bm{n})=0\,,
\end{align}
which, upon imposing the above mentioned constraint $\alpha_{ij}\geq 0$, gives:
\begin{align}
\alpha_{ij}=\frac{1}{2}\Bigg(1-\sigma^2_{ij}\mu_{ij}+\sqrt{1+4\sigma^2_{\alpha}n_{ij}-2\sigma^2_{\alpha}\mu_{ij}+\sigma^4_{\alpha}\mu^2_{ij}} \Bigg)\,,
\end{align}
which we substitute back into Eq.\eqref{LikelihoodFunction2} and Eq.\eqref{NegativeLogLikelihood}. Note that $\alpha_{ij}\rightarrow 1$ in the limit $\sigma_{\alpha}\rightarrow 0$. With the explicit dependence on $\bm{\alpha}$ eliminated from the likelihood function in Eq.\eqref{LikelihoodFunction2}, we may regard the nuisance parameters as the reduced set $\bm{\theta}=\{\bm{R}^\mathrm{CR},\bm{R}^\mathrm{GDE}\}$. At this stage we wish to find an upper limit of the 95$\%$ confidence interval for $\langle \sigma v \rangle$. Towards this end, we numerically calculate $\lambda$: first we find the denominator in Eq.\eqref{TS}, thereby obtaining $\bm{\mu}_{\text{MLE}}$; next, we modify the numerator by gradually increasing $\langle \sigma v \rangle$ until $\lambda$ reaches 2.71 (Recall that $\chi_1^{2} \sim Z^{2}$ with $Z\sim N(0,1)$, and hence for finding the upper limit of the 95\% confidence interval, we require $Z^{2}\approx 1.645^2 \approx 2.71$). We subsequently use this value of $\langle \sigma v \rangle$ for the 95\% upper limit in the exclusion line for one particular fixed mass $m_{F_{1}}$. We then repeat this method for masses ranging from 1 TeV to 100 TeV, thereby obtaining an exclusion line for $\langle \sigma v \rangle$.
\begin{figure}[h!]
\centerline{\includegraphics[width=12cm]{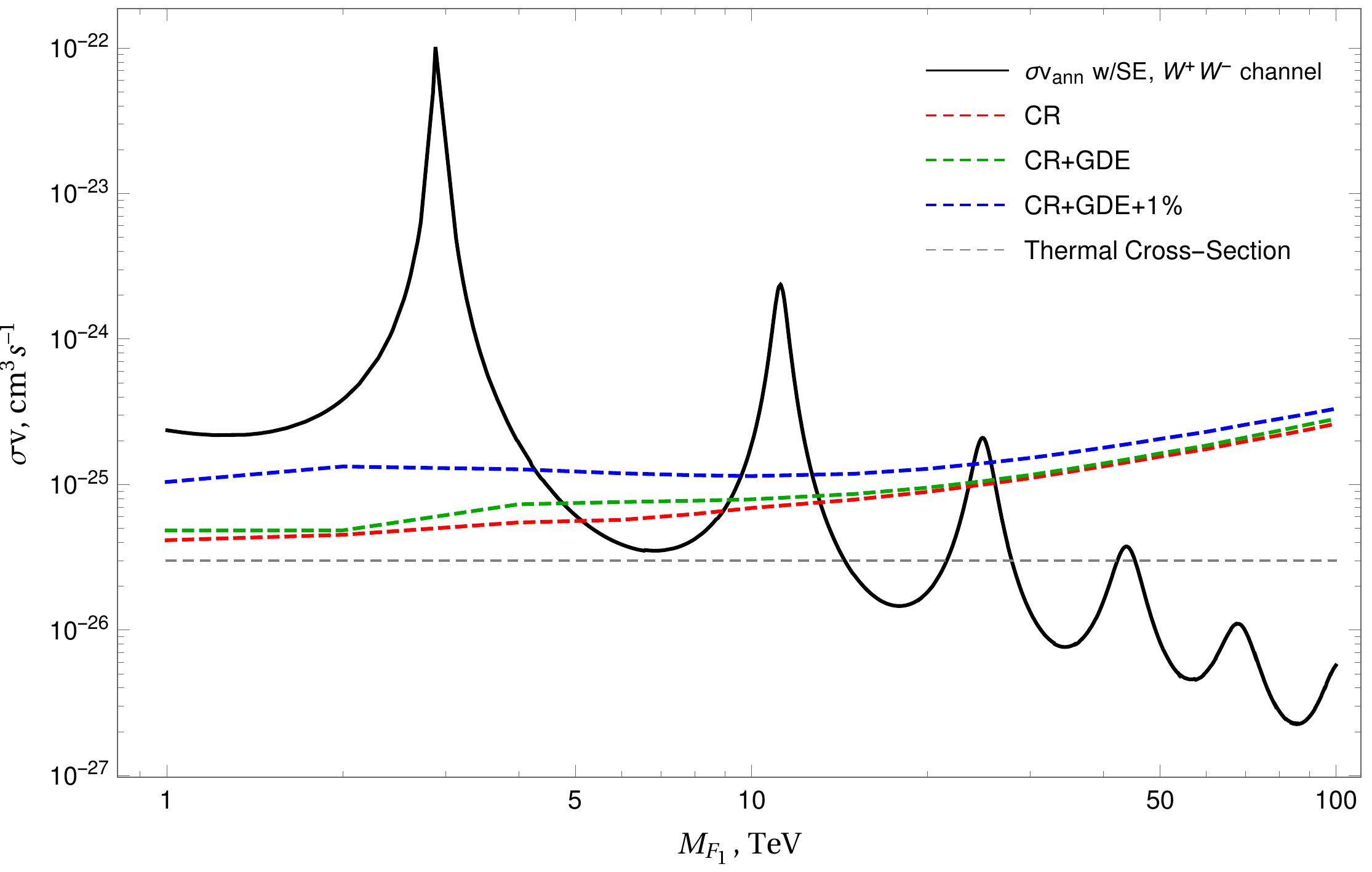}}
\caption{The projected upper limit on the annihilation cross-section for the triplet dark matter into $W^+ W^-$ final state for the NFW profile and 100 hour of observation at the CTA.}
\label{upperlimitfig}
\end{figure}
Finally, we repeat this entire process in the three distinct scenarios mentioned above, i.e. with the following backgrounds: CR, CR and GDE, as well as CR, GDE and 1\% systematic uncertainties. All of these results are illustrated in Fig. \ref{upperlimitfig}.

\subsection{Result and Discussion}\label{resultdiscussionsec}

In Fig. \ref{upperlimitfig}, we can see that the CTA can probe the DM of the triplet KNT model with the mass up to 25.7 TeV. The DM mass range $1.5\,\text{TeV}\leq M_{F_{1}}\leq 4.25\,\text{TeV}$, where the direct detection constraint sets the lower limit (Fig. \ref{dmrelic} (right)) , can be excluded the CTA observation of continuum gamma-rays coming DM annihilation at the GC. However, the CTA sensitivity derived here is subjected to several issues discussed in the following.

In our study, one prominent uncertainty which influences the expected gamma-ray count from the DM annihilation at the CTA and may degrade its DM detection sensitivity, is the astrophysical J-factor in Eq. \ref{flux1}. As the astrophysical J-factor depends on the DM density profile of the Milky Way,  the precise information of its density profile near the GC is essential to reduce the uncertainty that translates into the J-factor used for calculating gamma-ray flux. We have used the NFW profile which is cuspy at the GC to calculate the corresponding J-factors for our annular RoIs centering the GC. However, one could use joint profile likelihood analysis, as done, for example, in \cite{Rinchiuso:2018ajn}, to set upper limits on both the DM annihilation cross-section and the J-factor values associated with the considered RoIs.
 
Also,  the proper treatment of the background in RoIs contributes to the CTA sensitivity for DM detection. In our case, we consider the gamma-rays from the DM annihilation, and the backgrounds CR and GDE simultaneously from the same the region of the sky. One could have considered two regions of the sky, one where the count from DM signal is expected to be high, known as the ON region, and adjacent DM signal-free but background dominated region taken as OFF. The statistical significance of the DM signal is determined in this scenario by contrasting counts from ON and OFF region. However, in such ON-OFF method, it is not always possible to define the signal-free region beforehand for extended sources like the DM, which could lead to systematic uncertainties in the counts and eventually incorrect substantial significance. The method we use is less prone to systematic uncertainties as all the counts are taken for the same region.

Moreover, our extrapolation of the GDE up to 100 TeV from the Fermi P7V6 model given by Eq.(\ref{gdeeq}) is simplified as the GDE can have non-trivial spatial and spectral properties at such high energies. For this reason, we have derived CTA sensitivity with and without the GDE. If we consider only the CR as the background, we get the most stringent limit (red dashed line in Fig. \ref{upperlimitfig}) on the DM annihilation cross-section into $W^{+}W^{-}$ final states in the triplet KNT model at the GC from our analysis. The inclusion of the GDE reduces the sensitivity as seen from the green dashed line in Fig. \ref{upperlimitfig}. Finally, as expected, the systematics of $1\%$ further degrades the CTA sensitivity as illustrated by the blue dashed line in Fig. \ref{upperlimitfig}.

\section{Conclusion and Outlook}\label{conclusionsec}

In this work, we have studied the sensitivity of the upcoming Cherenkov Telescope Array to detect the DM candidate, with mass at the TeV range, of the KNT model annihilating into $W^{+}W^{-}$ at the Galactic Center. In the following, we summarize our key results in order.

As shown in section \ref{model}, the KNT model can be generalized with large fermionic multiplets, $\mathbf{F}_{i}$ of three generations and scalar multiplet, $\mathbf{\Phi}$, all of them having the same integer isospin, $j$ of $SU(2)_{L}$, and hypercharges, $Y_{F_{i}}=0$ and $Y_{\mathbf{\Phi}}=1$, respectively without changing the topology of three-loop neutrino mass generation.  Here, we set the upper limit on the value of the $SU(2)_{L}$ isospin for the KNT model using the bounds from the partial-wave unitarity on the coupled channels, $F^{(Q)}_{i}F^{(-Q)}\rightarrow W^{+}W^{-},\,ZZ,\,\gamma\gamma,\,\gamma\,Z$. The bound on the $SU(2)$ isospin of fermion multiplet is $j_{F}\leq 8$ for one generation and $j_{F}\leq 7$ for three generations, respectively. The inclusion of the elastic channels in the coupled-channel analysis makes the partial-wave unitarity bound somewhat ineffective because of the Coulomb singularity arising from photon exchange in $F_{i}^{(Q)}F_{i}^{(-Q)}\rightarrow F_{i}^{(Q)}F_{i}^{(-Q)}$.

Therefore, a more refined bound on the $SU(2)$ isospin comes from the appearance of the low-scale Landau pole in the gauge coupling in the presence of large electroweak multiplets. From Table \ref{landautab}, we can see that the $j=3$ case is almost ruled out and  $j=2$ is already in tension with the non-observation of non-perturbative $SU(2)$ coupling and the new physics appearing at the LHC. It leaves the minimal KNT with singlets and the triplet KNT with $j=1$ as the only viable models within the class of such three-loop radiative neutrino mass generation models.

However, the region of parameter space associated with the dark matter of the singlet and triplet KNT models are different. Because the DM of the singlet KNT does not have any SM gauge interaction, and its thermal freeze-out and annihilation at the present day are controlled by the KNT Yukawa couplings which directly connects them to the neutrino mass generation and the related low-energy constraints. For this reason, the viability of the singlet KNT DM with $O(\text{TeV})$ mass is left for a future study.  On the other hand, the DM of the triplet KNT model, being charged under the SM gauge group, can have large Sommerfeld enhancement, and its parameter space is determined mostly by the gauge interactions when the DM mass is in the TeV range. 

Besides, the SE annihilation of such heavy DM can produce detectable VHE gamma-rays at the IACTs. We show that the DM of the triplet KNT model annihilating into $W^{+}W^{-}$ at the GC can be probed up to 25.7 TeV by the future CTA experiment considering NFW profile and 100 hours of observation. However, there is some room for improvement of our sensitivity analysis. The triplet KNT model also has SE annihilation channels into $\gamma\gamma$ and $\gamma\,Z$ which would give line-like signature in the gamma-ray spectrum around, $E_{\gamma}\sim M_{F_{1}}$. Although the statistical analyses related to the continuum and line-like gamma-ray spectra have some differences, a proper combination of both analyses can lead to a better CTA sensitivity. Also, the consideration of the morphological analyses, adopted in \cite{Silverwood:2014yza, CTAConsortium:2018tzg, Acharyya:2020sbj} would result in an improved sensitivity of the CTA to detect the triplet KNT DM.

In addition, one can consider the dwarf spheroidal galaxies (dSphs), that is expected to contain up to O($10^{3}$) times more mass in DM than in visible matter and have low astrophysical gamma-ray background than the GC. Such characteristics of dSphs make them an ideal environment to search for the DM at the CTA \cite{CTAConsortium:2018tzg}. Furthermore, the electrons produced from DM annihilation (mostly as secondary) can generate synchrotron radiation in the presence of the magnetic field, which could be observed as a diffuse radio emission in the present and future radio observatories like Square Kilometer Array (SKA) \cite{Bull:2018lat}, and the sensitivity to detect such radio signal from DM in the dSphs is promising \cite{Colafrancesco:2015ola, Beck:2019ukt}. Therefore, a combined sensitivity study of the CTA and SKA  to detect the DM in the dSphs will enable us to probe further the parameter space of the generalized KNT model \cite{prep} and other variants \cite{Cepedello:2020lul,Gustafsson:2020bou} of the three-loop radiative neutrino mass generation models.

\section*{Acknowledgement}
We would like to thank Casas Bal\'{a}zs, Adil Jueid, J\"urgen Kn\"odlseder, Luca Di Luzio, Ernest Ma, Soebur Razzaque, Nicholas L. Rodd and Piero Ullio for stimulating discussions. TAC and SN would like to thank The Abdus Salam International Centre for Theoretical Physics for the hospitality and support where part of the work was done. SH is supported by the Prime Minister Fellowship, Prime Minister's Office, Government of Bangladesh. This research has made use of the CTA instrument response functions provided by the CTA Consortium and Observatory, see http://www.cta-observatory.org/science/cta-performance/ (version prod3b-v2) for more details. Also, this research made use of ctools, a community-developed analysis package for Imaging Air Cherenkov Telescope data. ctools is based on GammaLib, a community-developed toolbox for the scientific analysis of astronomical gamma-ray data. 

\appendix

\section{Helicity Amplitudes and Partial-wave Unitarity}\label{partialapp}
The helicity eigenstates of the fermion ($u_{\pm}$) and antifermion ($v_{\pm}$) with respect to the following 3-momentum,
\begin{equation}
\vec{p}=(p\sin\theta\, \cos\phi,\,p\sin\theta\,\sin\phi,\,p\cos\theta)\,,
\label{par1}
\end{equation}
are given by,
\begin{align}
u_{+}(\vec{p})&=\sqrt{E+m}\left(\cos\frac{\theta}{2},\,e^{i\phi}\sin\frac{\theta}{2},\,\frac{p}{E+m}\cos\frac{\theta}{2},\,\frac{p}{E+m}e^{i\phi}\sin\frac{\theta}{2}\right)\,,\label{par2}\\
u_{-}(\vec{p})&=\sqrt{E+m}\left(-\sin\frac{\theta}{2},\,e^{i\phi}\cos\frac{\theta}{2},\,\frac{p}{E+m}\sin\frac{\theta}{2},\,-\frac{p}{E+m}e^{i\phi}\cos\frac{\theta}{2}\right)\,,\label{par3}\\
v_{+}(\vec{p})&=\sqrt{E+m}\left(\frac{p}{E+m}\sin\frac{\theta}{2},\,-\frac{p}{E+m}e^{i\phi}\cos\frac{\theta}{2},\,-\sin\frac{\theta}{2},\,e^{i\phi}\cos\frac{\theta}{2}\right)\,,\label{par4}\\
v_{-}(\vec{p})&=\sqrt{E+m}\left(\frac{p}{E+m}\cos\frac{\theta}{2},\,\frac{p}{E+m}e^{i\phi}\sin\frac{\theta}{2},\,\cos\frac{\theta}{2},\,e^{i\phi}\sin\frac{\theta}{2}\right)\,,\label{par5}
\end{align}
where $\pm$ denotes either along or opposite to the given momentum direction. The incoming momenta of fermion and anti-fermion are taken  along the $z$-axis ($\theta=0,\,\phi=0$) and opposite to it ($\theta=\pi,\,\phi=\pi$), respectively.

The transverse (T) and longitudianl (L) polarization 4-vectors of the outgoing gauge bosons with respect to the 3-momentum, $\vec{p}$ is given by,
\begin{align}
\epsilon_{T}(\hat{p},\lambda)&=\frac{1}{\sqrt{2}}\left(0,\,-\lambda \cos\theta\cos\phi+i\sin\phi,\,-\lambda\cos\theta\sin\phi-i\cos\phi,\,\lambda\sin\theta\right),\,\,\lambda=\pm 1\,,\label{pol1}\\
\epsilon_{L}(\hat{p})&=\frac{1}{M_{V}}\left(|\vec{p}|,\,E \sin\theta\cos\phi,\,E\sin\theta\sin\phi,\,E\cos\theta\right)\,.\label{pol2}
\end{align}
Therefore, if the outgoing gauge bosons are in the $xz$ plane, their corresponding polarization 4-vectors are given by setting $(\theta,\phi=0)$ and $(\pi-\theta,\phi=\pi)$, respectively.

In the high energy limit, $\sqrt{s}\rightarrow\infty$, the helicity conserving amplitude ($|\mu_{i}|=|\mu_{f}|$) can be expanded as,
\begin{equation}
\mathcal{M}=\mathcal{M}_{\frac{1}{2}}s^{\frac{1}{2}}+O(1/\sqrt{s})\,,
\label{pol3}
\end{equation}
whereas the helicity violating amplitude ($|\mu_{i}|\neq|\mu_{f}|$) can be expanded as,
\begin{equation}
\mathcal{M}=\mathcal{M}_{1}s+\mathcal{M}_{0}+O(1/s)\,.
\label{pol4}
\end{equation}
As we are taking into account the high-energy tree-level scattering, the above-mentioned helicity amplitudes between the initial states, $|i\rangle=|F^{(Q)}_{i}F^{(-Q)}_{i}\rangle$ and final states, $|f\rangle=|V\,V^\prime\rangle$ (with $V\,V^{\prime}=W^{+}W^{-},ZZ,\gamma\gamma,\gamma Z$), are considered to be the matrix elements ${\cal T}_{fi}$ of Eq.(\ref{partial1}) and Eq.(\ref{partial2}). In the subsequent paragraphs,  we avoid writing the values of subscripts $i$ and $f$ in ${\cal M}$ explicitly to have simplified notations.

We have checked that the cancellations among s-channel, t-channel and u-channel in $FF\rightarrow VV$ scattering render the terms, $\mathcal{M}_{\frac{1}{2}}$ and $\mathcal{M}_{1}$ to be zero which is expected as $\mathbf{F}_{i}$ are the electroweak multiplets. In the following we tabulate only the $\mathcal{M}_{0}$ terms of the corresponding helicity amplitudes.

For $F^{(Q)}F^{(-Q)}\rightarrow W^{+}W^{-}$ where $W^{\pm}$ are transverse, we have,
\begin{eqnarray}
\mathcal{M}_{\mathrm{0}}(+-;+-) & = & g^{2}\left[2\tan\frac{\theta}{2}\left\{V_{-}(t_{3})V_{+}(t_{3}-1)\right\}-t_{3}\sin\theta\right]\,,\label{app1}\\
\mathcal{M}_{\mathrm{0}}(-+;+-) & = & -g^{2}\left[2\cot\frac{\theta}{2}\left\{V_{+}(t_{3})V_{-}(t_{3}+1)\right\}+t_{3}\sin\theta\right]\,,\label{app2}
\end{eqnarray}
where, the first two entries correspond to fermion's helicity, $\pm\frac{1}{2}$ and last two entries denote gauge boson's helicity, $\pm 1$. Also, $t_{3}$ is the eigenvalue of the diagonal generator, $T^{3}$ of $SU(2)$ associated with the component field, $F^{(Q)}_{i}$ of the multiplet, $\mathbf{F}_{i}$, and $t_{3}$ takes value as, $t_{3}=-j_{F},\,-j_{F}+1,...,j_{F}-1,\,j_{F}$, if $j_{F}$ is the isospin of the fermionic multiplet. Besides, as the hypercharge of the fermion multiplet is zero, the electric charge of its component field follows, $Q=t_{3}$. Moreover, $V_{\pm}(t_{3})=\frac{1}{\sqrt{2}}\sqrt{(t\mp t_{3})(t\pm t_{3}+1)}$ corresponds to $SU(2)$ raising and lowering factors with isospin, $t$ and $t_{3}$.

In addition, for the scattering $F^{(Q)}F^{(-Q)}\rightarrow ZZ$ where $Z$ is transverse, we have,
\begin{eqnarray}
\mathcal{M}_{0}(+-;+-)=\mathcal{M}_{0}(-+;-+) & = &\sqrt{2}g^{2}t_{3}^{2}\cos^{2}\theta_{w}\tan\frac{\theta}{2}\,,\label{app3}\\
\mathcal{M}_{0}(+-;-+)=\mathcal{M}_{0}(-+;+-) & = & -\sqrt{2}g^{2}t_{3}^{2}\cos^{2}\theta_{w}\cot\frac{\theta}{2}\,.\label{app4}
\end{eqnarray}

Also, for $F^{(Q)}F^{(-Q)}\rightarrow \gamma\gamma$ scattering,
\begin{eqnarray}
\mathcal{M}_{0}(+-;+-)=\mathcal{M}_{0}(-+;-+) & = &\sqrt{2}g^{2}t_{3}^{2}\sin^{2}\theta_{w}\tan\frac{\theta}{2}\,,\\
\mathcal{M}_{0}(+-;-+)=\mathcal{M}_{0}(-+;+-) & = & -\sqrt{2}g^{2}t_{3}^{2}\sin^{2}\theta_{w}\cot\frac{\theta}{2}\,.
\end{eqnarray}

Finally, we have for $F^{(Q)}F^{(-Q)}\rightarrow \gamma Z$, where $Z$ is transverse,
\begin{eqnarray}
\mathcal{M}_{0}(+-;+-)=\mathcal{M}_{0}(-+;-+) & = &2g^{2}t_{3}^{2}\sin\theta_{w}\cos\theta_{w}\tan\frac{\theta}{2}\,,\label{app5}\\
\mathcal{M}_{0}(+-;-+)=\mathcal{M}_{0}(-+;+-) & = & -2g^{2}t_{3}^{2}\sin\theta_{w}\cos\theta_{w}\cot\frac{\theta}{2}\,.\label{app6}
\end{eqnarray}

Moreover, as we can see from Eqs.(\ref{app1})-(\ref{app6}) that the helicity amplitudes have initial total helicity either as $\mu_i=1$ or $\mu_{i}=-1$, and final total helicity either as $\mu_f=2$ or $\mu_f=-2$. Therefore the Wigner d-matrices of Eq.(\ref{partial2}) will involve, $d^{J}_{{\pm 1},{\pm 2}}$ with $J=2,3,..$, corresponding to these helicity amplitudes. Since typically partial wave amplitudes with higher $J$ values are smaller compared to those with lower possible $J$ values, we consider the partial wave amplitudes with $J=2$ in our analysis. Hence, we have used the following Wigner d-matrices to calculate $a^{J=2}_{fi}$.
\begin{align}
d_{2,1}^{2} & = -d_{1,2}^{2}= -\frac{1}{2}(1+\cos\theta)\sin\theta\,,\nonumber\\
d_{2,-1}^{2} & = -d_{-1,2}^{2}=-\frac{1}{2}\sin\theta(1-\cos\theta)\,.\nonumber
\end{align}

The coupled-channel matrix is written the following basis,
\begin{equation}
|\psi_{sc}\rangle=\left(W^{+}W^{-},ZZ,\gamma\gamma,\gamma\,Z,F_{i}^{0}F^{0}_{i},..,F^{(Q)}_{i}F^{(-Q)}_{i},..,F^{(j_{F})}_{i}F^{(-j_{F})}_{i}\right)^{T}\,,\label{pol5}
\end{equation}
where, $i=1,2,3$ is the generation index and for more than one generations of multiplets, the 2-particle states $F^{(Q)}_{i}F^{(-Q)}_{i}$ are repeated for each value of $i$ in the above basis. For example, for isospin, $j_{F}=1$ and one generation, it is given by,
\begin{equation}
a^{J=2}=\begin{pmatrix}
0 & 0 & 0 & 0 & \frac{g^2}{24\sqrt{2}\pi} & \frac{g^2}{48\pi} \\
0 & 0 & 0 & 0 & 0 & \frac{g^2\cos^{2}\theta_{w}}{24\sqrt{2}\pi} \\
0 & 0 & 0 & 0 & 0 & \frac{g^2\sin^{2}\theta_{w}}{24\sqrt{2}\pi} \\
0 & 0 & 0 & 0 & 0 & \frac{g^2\cos\theta_{w}\sin\theta_{w}}{24\pi} \\
\frac{g^2}{24\sqrt{2}\pi} & 0 & 0 & 0 & 0 & 0\\
\frac{g^2}{48\pi} & \frac{g^2\cos^{2}\theta_{w}}{24\sqrt{2}\pi} & \frac{g^2\sin^{2}\theta_{w}}{24\sqrt{2}\pi} & \frac{g^2\cos\theta_{w}\sin\theta_{w}}{24\pi} & 0 & 0
\end{pmatrix}\,.
\label{app7}
\end{equation}
As we are only considering $FF\rightarrow VV^\prime$ scattering, the matrix elements of Eq.(\ref{app7}) are zero for $V\,V^{\prime}\rightarrow V\, V^{\prime}$ and $FF\rightarrow FF$ channels. As the order of the coupled-channel matrix increases when the isospin of the fermion multiplet, $j_{F}$ and generation number, $i$ increase, we first determine the largest eigenvalue of the corresponding coupled-channel matrix and identify the largest values of $j_{F}$ which satisfy the bound in Eq.(\ref{partial7}) for single and three generations, respectively.

In  the case of high energy scattering of $F^{(Q)}F^{(-Q)}\rightarrow F^{(Q)}F^{(-Q)}$ with $Q\neq 0$, the photon exchange at the $t$-channel would lead to the following helicity amplitudes,
\begin{align}
{\cal M}^{(t)}_{0}(+\frac{1}{2},+\frac{1}{2};+\frac{1}{2},+\frac{1}{2})&=2Q^{2}e^{2}\,\mathrm{cosec}^{2}(\theta/2),\,\,\,(\mu_{i}=0,\,\mu_f=0)\,,\label{app8}\\
{\cal M}^{(t)}_{0}(+\frac{1}{2},-\frac{1}{2};+\frac{1}{2},-\frac{1}{2})&=2Q^{2}e^{2}\,\mathrm{cot}^{2}(\theta/2),\,\,\,(\mu_{i}=1,\,\mu_f=1)\,.\label{app9}
\end{align}
From Eq.(\ref{app8}) and (\ref{app9}), we can see that only Wigner d-matrices, $d^{J}_{00}(\theta)$ and $d^{J}_{\pm 1,\pm 1}(\theta)$ will be present in Eq.(\ref{partial2}) but then the integration over $\cos\theta$ will be divergernt because of the integrand's singularities at $cos\theta=\pm 1$. This is the Coloumb singularity appearing in the elastic channels, which is mentioned in section \ref{partialsec}.

\section{Landau Pole}\label{landauapp}

The one-loop beta-function for the $SU(2)$ coupling given in Eq.(\ref{lan1}) is,
\begin{equation}
\beta_{g}=\frac{g^{3}}{16\pi^{2}}\left(-\frac{19}{6}+\frac{4}{3}\sum_{i}\kappa_{F_{i}}T(F_{i})+\frac{1}{3}\sum_{j}\eta_{S_{j}}T(S_{j})\right)=\frac{g^{3}}{16\pi^{2}} b_{g}\,,
\label{lanapp1}
\end{equation}
where, $b_{g}$ is denoting the factor of $g^{3}/16\pi^2$ in Eq.(\ref{lanapp1}). Therefore, the coupling $g(\mu)$ at the energy scale, $\mu$ is given as,
\begin{equation}
g(\mu)=\frac{g(\mu_0)}{1-\frac{b_{g} g(\mu_{0})^{2}}{8\pi^{2}}\mathrm{ln}\left(\frac{\mu}{\mu_0}\right)}\,,
\label{lanapp2}
\end{equation}
where, $\mu_{0}$ is the reference energy scale and $g(\mu_0)$ is the value of coupling at that scale. Moreover, we denote $b^{\mathrm{SM}}_{g}=-19/6$ for the SM contribution only. Now, if $b_{g}$ is positive, we have an energy scale called the Landau pole, $\mu=\Lambda_{\mathrm{Lan}} > \mu_0$, for which the denominator of Eq.(\ref{lanapp2}) vanishes (provided $g(\mu_0)\neq 0$), and we get,
\begin{equation}
\Lambda_{\mathrm{Lan}}=\mu_0\, \mathrm{exp}\, \left(\frac{8\pi^2}{b_{g}\,g(\mu_{0})^{2}}\right)\,.
\label{lanapp3}
\end{equation}
Now, for the generalized KNT model, there are three generations of $SU(2)$ fermion multiplets with isospin $j$ (where, $j$ is integer) and hypercharge, $Y=0$, and one $SU(2)$ scalar multiplet with isospin $j$ and hypercharge, $Y=1$. Therefore, the factor $b_{g}$ is given by, $b_{g}=-\frac{19}{6}+\frac{13}{9}j(j+1)(2j+1)$. Consequently, the factor, $b_{g}$ becomes positive for isospin, $j\geq 1$. We consider the electroweak (EW) scale at $\mu_{\mathrm{EW}}=M_{Z}=91.1876$ GeV with $g(M_{Z})=0.65114$ \cite{DiLuzio:2015oha, PDG}. If the masses of the additional KNT multiplets, which are, for simplicity, denoted by a common mass parameter, $M_{X}$,  are comparable to the electroweak scale, then the running of the $SU(2)$ coupling from the electroweak scale, $\mu_0=M_{Z}$ to the higher energy will involve positive $b_{g}$ for isospin $j\geq 1$, and correspondingly we determine the Landau pole of $g$ using Eq.(\ref{lanapp3}) which are tabulated in the second column of Table \ref{landautab} for $j=1,2,3$. Now, if the KNT mass-scale is $M_{X}\gg M_{Z}$, i.e. in the TeV range, the contributions from the KNT fields to the running of the $SU(2)$ from $M_{Z}$ to $M_{X}$ can be considered negligible. In this case, we first determine, $g(\mu=M_{X})$ using Eq.(\ref{lanapp1}) with $b_{g}=b^{\mathrm{SM}}_{g}$ from the EW scale, $\mu_{0}=M_{Z}$, that is up to the energy scale, $M_{X}$ where the KNT fields' contributions in the running become relevant. Afterwards, we compute the Landau pole for isospin $j$ using Eq.(\ref{lanapp3}) with $\mu_{0}=M_{X}$ and $g(\mu_{0}=M_{X})$ as inputs. We tabulate the corresponding Landau poles for the isospin, $j=1,2,3$ in the third, fourth and fifth columns of Table \ref{landautab} for $M_{X}=10^{3},\,10^{4},\,10^{5}$ GeVs, respectively.

\end{document}